\documentclass[12pt,a4paper]{article}

\usepackage{amsthm}
\usepackage{amsmath}
\usepackage{amssymb}
\usepackage{natbib}
\usepackage[colorlinks,citecolor=blue,urlcolor=blue,filecolor=blue,backref=page]{hyperref}
\usepackage{graphicx}

\usepackage{multirow,bigdelim}

\newcommand{ \uv}{{\boldsymbol{u}}}
\newcommand{\Rv}{{\mathbf{R}}}
\newcommand{\Av}{{\mathbf{A}}}
\newcommand{\bv}{{\mathbf{b}}}
\newcommand{\ev}{{\mathbf{e}}}
\newcommand{\Bv}{{\mathbf{B}}}

\newcommand{\xv}{{\mathbf{x}}}
\newcommand{\Xv}{{\mathbf{X}}}
\newcommand{\yv}{{\mathbf{y}}}
\newcommand{\zv}{{\mathbf{z}}}
\newcommand{\qv}{{\mathbf{q}}}
\newcommand{\Iv}{{\mathbf{I}}}
\newcommand{ \Uv}{{\mathbf{U}}}

\newcommand{\zerov}{{\mathbf{0}}}
\newcommand{\Dv}{{\mathbf{D}}}

\newcommand{\Qv}{{\mathbf{Q}}}
\newcommand{\Wv}{{\mathbf{W}}}
\newcommand{\Zv}{{\mathbf{Z}}}

\newcommand{\errorv}{\boldsymbol{\varepsilon}}

\newcommand{\alphav}{\boldsymbol{\alpha}}
\newcommand{\betav}{\boldsymbol{\beta}}
\newcommand{\deltav}{\boldsymbol{\delta}}

\newcommand{\kappav}{\boldsymbol{\kappa}}

\newcommand{\tauv}{\boldsymbol{\tau}}
\newcommand{\thetav}{\boldsymbol{\theta}}

\newcommand{\zetav}{\boldsymbol{\zeta}}

\newcommand{\Sigmav}{\boldsymbol{\Sigma}}
\newcommand{\Deltav}{\boldsymbol{\Delta}}
\newcommand{\Omegav}{\boldsymbol{\Omega}}

\newcommand{\Normal}[1]{ \mathcal{N}\left(#1\right)}

\newcommand{\Gammainv}[1]{\mathcal{G}^{-1}\left(#1\right)}
\newcommand{\Exp}[1]{ \mathcal{E}\left(#1\right)}
\newcommand{\Gammad}[1]{\mathcal{G}\left(#1\right)}

\DeclareMathOperator{\E}{E}

\DeclareMathOperator{\Cov}{Cov}
\DeclareMathOperator{\Prec}{Prec}
\DeclareMathOperator{\Corr}{Cor}
\DeclareMathOperator{\diag}{diag}

\begin{document}


\title{Bayesian Effect Fusion for Categorical Predictors}

\author{Daniela Pauger, Helga Wagner}

\maketitle
\begin{abstract}
We propose a Bayesian approach to  obtain a sparse representation  of the effect of a categorical predictor in regression type models. As this effect  is captured by a group of level effects, sparsity cannot only be achieved by excluding single irrelevant level effects or  the whole group of effects associated to this predictor  but also by fusing levels which have essentially the same effect on the response. To achieve this goal, we propose a prior which allows for almost perfect as well as almost zero dependence between level effects a priori. This prior can alternatively be obtained by specifying spike and slab prior distributions on all effect differences associated to this categorical predictor. We show how  restricted  fusion can be implemented and develop an efficient MCMC method  for posterior computation. The performance of the proposed  method is investigated  on simulated data and we illustrate its application on real data from EU-SILC.  
\end{abstract}

\vspace{1.5cm} 
\emph{keywords}: spike and slab prior, sparsity, nominal and ordinal predictor,  regression model, MCMC, Gibbs sampler
\vspace{1.5cm}

\thispagestyle{empty}

\section{Introduction}

In many applications, especially in medical, social or economic studies, potential covariates collected  for a  regression analysis are categorical, measured either on an ordinal or on a nominal scale. The usual strategy for modelling the effect of a categorical covariate is to define one level as baseline and to  use dummy variables for the effects of the other levels with respect to this baseline. Hence, the effect of  a categorical covariate  is not captured   by a single  but by  a  group of regression effects. Including  categorical variables as covariates in regression type models can therefore easily lead to a high-dimensional vector of regression effects.  Moreover,  as only observations with a specific level contribute
information on this level effect, estimated effects of rare levels will be associated with high uncertainty.  

Many methods have been proposed  to  achieve sparsity in regression models by identifying regressors with non-zero effects. Whereas frequentist methods, e.g.~the lasso \citep{tib:reg} or the elastic net \citep{zou-has:reg}  rely on  penalties,  Bayesian variable selection methods are based on the specification of appropriate prior distributions, e.g.~shrinkage priors \citep{par-cas:bay, gri-bro:inf} or spike and slab priors  \citep{mit-bea:bay, geo-mcc:app, ish-rao:spi}. However, variable selection methods  identify  single non-zero regression effects  and   do not take into account the natural grouping of the set of  dummy variables capturing the effect of a categorical covariate.
 
Moreover, for a categorical covariate  a sparser representation of its effect cannot only be achieved  by  restricting some or all  of its level effects to  zero but also when some of the  levels have the same effect. To address this problem, we propose a Bayesian approach  to achieve a sparsity  by  encouraging  both shrinkage of non-relevant effects to zero as  well as fusion of (almost) identical  level effects.  
   
Methods that explicitly  address inclusion  or exclusion of  a  whole group of regression coefficients are  the group lasso \citep{yua-lin:mod}, the Bayesian  group lasso \citep{ram-etal:bay, kyu-etal:pen} and the approach  of \cite{chi:bay}  who uses spike and slab priors for grouped selection of the set of  dummy  variables related to a categorical predictor. The recently proposed sparse group lasso \citep{sim-etal:spa}  and  Bayesian sparse group selection   \citep{che-etal:bay}  aim at  sparsity  at the group level as well as   within selected groups by  selecting non-zero regression effects, but do not consider fusion of effects.

For metric predictors,  effect fusion is addressed in  \cite{tib-etal:spa}  with  the fused lasso   and  in  \cite{kyu-etal:pen} with  its Bayesian counterpart,   the Bayesian fused lasso.   Both methods assume some ordering of effects and shrink only effect differences of subsequent effects to zero.  Hence, they  are  not appropriate  for nominal predictors where any effect difference should be subject to shrinkage. 

 Sofar, only a few papers  consider effect fusion for nominal predictors.       \cite{bon-rei:sim}  propose a modification of the fused lasso for  ANOVA and  Gertheiss and Tutz \citep{ger-tut:pen, ger-tut:spa, tut-ger:reg}   specify different lasso-type penalties for ordinal and nominal covariates. Recently,  \cite{tut-ber:tre} propose  tree-structured clustering of effects of categorical covariates. In a Bayesian approach,  fusion (or merging) of levels of  categorical variables 
is addressed only in \cite{del-tar:mod}. Their  goal is  to analyse dependence of categorical variables in loglinear models. To search the  huge  space of models, that can be obtained by collapsing  levels of categorical variables, a reversible jump algorithm is employed. This 
method could be  extended   to search the space of regression type models where levels of categorical predictors are subject to merging, but  we suggest  a different approach where Bayesian inference is feasible via a simple Gibbs sampling  algorithm. 

To allow for effect fusion we  specify a joint multivariate  Normal prior for all level effects of one covariate with  a precision matrix  that allows for either almost perfect or low dependence  of regression effects. This prior is related to spike and slab prior distributions that have been  applied  extensively in Bayesian approaches to variable selection. We show that the  prior proposed for effect fusion can be derived alternatively by specifying spike and slab prior distributions on all  level effects as well as  their differences and taking into account  their linear dependence. Whereas with the usual variable selection prior   regression effects  can be classified as (almost) zero, if  assigned to the spike, and as non-zero otherwise, the effect fusion prior allows for intrinsic classification  of effects as well as of  effect differences as negligible or relevant.   In contrast to  a categorical predictor where  typically each pair of  level effects will be subject to fusion,  for an ordinal predictor the available ordering information  can be exploited  by restricting fusion to adjacent categories. We show how  restriction  of effect fusion  to specific pairs of effects can be implemented in our framework.

For ease of exposition we will discuss construction of the prior and MCMC inference for a Normal  linear regression model with categorical covariates. However, the  method we propose can be  used  in  any regression type model  with a structured additive predictor, where additionally to effects of  categorical covariates further effects,  e.g.    
   linear or nonlinear effects of continuous covariates,   spatial  or random  effects are included.

The rest of the paper is organised as follows: in  Section \ref{sec:model} we introduce  the  data model and   in Section \ref{sec:prior} we describe construction of  the prior distribution to encourage a  sparse representation of the  effect of a categorical  predictor. Posterior inference is discussed in Section \ref{sec:inference} and Section \ref{sec:simulation} investigates the performance of the method for simulated data. Application of  Bayesian effect fusion is illustrated on a real data example in Section \ref{sec:application} and we  conclude with Section \ref{sec:conclusion}.

\section{Model specification} \label{sec:model}

We consider a standard linear regression model with Normal response   $y$ and  $p$ categorical covariates  where covariate $h$ has $c_h+1$ ordered or unordered  levels
 $0,\dots, c_h$. To represent the  effect of covariate $h$ on the response $y$, we define $0$ as the baseline category and introduce dummy variables,  $X_{h,k}$,  to capture the effect  of level $k$. The regression  model is  then given as 
 \begin{align}
	y= \mu + \sum_{h=1}^p \sum_{k=1}^{c_h}  X_{h,k} \beta_{h,k}+ \varepsilon,
	\label{regmod1}
\end{align}
where $\mu$ is the intercept,    $\beta_{h,k}$ is  the effect of level  $k$   of  covariate $h$ (with respect to the baseline category $0$) and  $\varepsilon \sim \Normal{0,\sigma^2}$ is the  error term.  

For  an $(n \times 1)$   response vector $\yv=(y_1,\dots, y_n)'$   we write  the model as 
\begin{equation}  \yv= \mathbf{1} \mu + \sum_{h=1}^p \Xv_h \betav_h +\boldsymbol{\varepsilon}, \quad    \boldsymbol{\varepsilon} \sim \Normal{\mathbf{0},\sigma^2 \mathbf{I}}, \label{model1} \end{equation}
where  $\Xv_h$  is the $(n \times c_h)$ design matrix for   covariate $h$, $\betav_h$  is the  $(c_h \times 1)$ vector of  the corresponding regression effects and $\boldsymbol{\varepsilon}$  the $(n \times 1)$ vector of error terms. $\mathbf{1}$ denotes a vector  with elements $1$ and   $\mathbf{I}$   the identity matrix.

\section{Prior specification} \label{sec:prior}

Bayesian model specification is completed by assigning  prior distributions  to all model parameters. We assume  a prior of the structure
$$p(\mu,\betav_1,\dots, \betav_p,\sigma^2)=p(\sigma^2)  p(\mu)\prod_{h=1}^p p(\betav_h|\tau^2_h,\deltav_h)p(\tau^2_h)p(\deltav_h),$$
where $\tau^2_h$ and $\deltav_h$ denote additional hyperparameters, which are specified below. We assign  a flat proper  prior $p(\mu) \sim \Normal{0,M_{0}}$ to the intercept,  and an Inverse Gamma prior $p(\sigma^2) \sim \Gammainv{s_0, S_0}$ to the error variance.
In our analyses we use the standard  improper prior  $p(\sigma^2) \propto 1 /\sigma^2$. 
 
To allow for effect fusion we specify the prior on the regression effects $\betav_h$ hierarchically as 
\begin{align}  \betav_h|\tau^2_h,\deltav_h   &  \sim \Normal{\zerov,\Bv_{h0}(\deltav_h, \tau^2_h)}, \label{pri1}\\
\tau_h^2 & \sim \mathcal{G}^{-1}(g_{h0}, G_{h0}), \label{prior_tau} 
\end{align}
with  prior covariance matrix of $\betav_h$  given as \begin{equation} \Bv_{h0}(\deltav_h,\tau^2_h)= \gamma_h \tau_h^2 \Qv_h^{-1}(\deltav_h).\label{privar}
\end{equation}
Here  $\gamma_h$ is a fixed constant, $\tau_h^2$ is a scale parameter and  the  matrix $\Qv_h(\deltav_h)$ determines the structure of the prior precision matrix of $\betav_h$. To encourage effect fusion,  we let  $\Qv_h(\deltav_h)$  depend on a vector $\deltav_h$ of  binary indicator variables $\delta_{h,kj}$, which are defined for each pair of level effects $\beta_{h,k}$ and $\beta_{h,j}$ subject to fusion.  $\delta_{h,kj}=1$ indicates that    $\beta_{h,k}$ and $\beta_{h,j}$ differ considerable and hence two regression parameters are needed to capture their respective effects whereas for  $\delta_{h,kj}=0$ the effects are almost identical and the two level effects could be fused. To allow also fusion of level effects to $0$, i.e.~conventional  variable selection, we  define $\beta_{h,0}=0$ and  include in $\deltav_h$ also indicators  $\delta_{h,k0}$, $k=1,\dots c_h$.

The dimension of $\deltav_h$   and  the  concrete specification of  $\Qv_h(\deltav_h)$ depend on  which pairs  of effects are subject to fusion.
For a  nominal covariate  typically levels are completely unstructured and hence any pair of effects might be fused.  We discuss this case where fusion is unrestricted  in Section \ref{sec:pri_nom}. 

  In contrast,  for an ordinal covariate the information on the ordering of levels suggests  to restrict fusion to  adjacent categories \citep{ger-tut:pen}.  Restrictions that preclude direct fusion for specified pairs of effects can easily be 
 implemented in the specification of the prior covariance matrix $\Bv_{h0}(\deltav_h,\tau^2_h)$. We present two different ways to incorporate fusion restrictions in Section \ref{sec:pri_restr}  and discuss the case of an ordinal covariate  where  fusion is  restricted to adjacent categories in more detail.

Marginalized over the indicators $\deltav_h$, the prior on $\betav_h$ is a mixture of multivariate Normal distributions 
 $$p( \betav_h|\tau^2_h)= \sum_{\deltav_h} p(\deltav_h) f_\mathcal{N}\big(\betav| \zerov, \Bv_{h0}(\deltav_h, \tau_h^2) \big) ,$$
 with  different component covariance matrices depending  on  $\deltav_h$ and   mixture weights $p(\deltav_h)$.
  We discuss specification  of the prior on  $\deltav_h$  in Section \ref{sec:pri_ind}  and  the choice of the 
 hyperparameters in  Section \ref{sec:hyp}.
 
For notational convenience we drop  the covariate index $h$ in the rest of this Section.

\subsection{Prior for unrestricted effect fusion }\label{sec:pri_nom}
To perform  unrestricted  effect fusion for a categorical covariate with levels $0,\dots, c$,  we introduce  a   binary  indicator $\delta_{kj}$  for each pair of effects   $0 \le j < k \le c$. Thus, the  vector  $\deltav$  subsuming  all these indicators is of dimension $d \times 1$ where  $d={\binom{c+1} {2}}$.  

We specify the structure  matrix $\Qv(\deltav)$  as  
\begin{equation}\label{eq:prec}
\Qv(\deltav)=\begin{pmatrix}  \sum_{j \neq 1} \kappa_{1j} & -\kappa_{12}  & \dots & -\kappa_{1c}\\
-\kappa_{21} &   \sum_{j \neq 2} \kappa_{2j} & \cdots  & -\kappa_{2c}\\
\vdots & \vdots & \ddots & \vdots\\
-\kappa_{c1} & -\kappa_{c2} & \dots & \sum_{j \neq c} \kappa_{cj} \end{pmatrix} 
\end{equation}
with diagonal elements $q_{kk}$ given as 
$$q_{kk}=\sum_{j \neq k} \kappa_{kj}=  \kappa_{k0}+\dots+ \kappa_{k,j-1}+\kappa_{k,j+1}+\kappa_{k,c} \qquad k=1,\dots, c. $$ 
For  $k >j$, $\kappa_{kj}$ is defined as
 $$\kappa_{kj}=\delta_{kj}+ r (1-\delta_{kj}), $$
 and   $\kappa_{jk}=\kappa_{kj}$  for $j>k$.
 The value of  $\delta_{kj}$ determines whether  $\kappa_{kj}=1$ (for $\delta_{kj}=1$) or $\kappa_{kj}=r$ (for $\delta_{kj}=0$).  $r$ is a fixed large number, which we call precision ratio for reasons explained below.
  Finally,  we set  $\gamma=c/2$.  

We  discuss this specification now in more detail. First, as  the structure matrix $\Qv(\deltav)$ determines  the prior precision matrix    $\Bv_0^{-1}(\deltav, \tau^2)$ up to the scale factor $1/(\gamma \tau^2)$  it 
has to be symmetric and positive definite. Symmetry  of $\Qv(\deltav)$ is guaranteed by definition and positive definiteness as 
 \begin{equation}
 \betav' \Qv (\deltav) \betav= \sum_{k =1}^c \beta_k^2  \kappa_{k0} + \sum_{k =2}^c  \sum_{j=1} ^{k-1}
  (\beta_k - \beta_j)^2   \kappa_{kj}>0,\label{eq:quadr_form}
\end{equation}
if  $\betav \neq \zerov$, see Appendix \ref{app:pri_pd} for  a  detailed proof. 

The  diagonal elements $q_{kk}$ determine the prior partial  precisions  and the off-diagonal elements  $q_{kj}$ the prior partial  correlations of the level effects:
\begin{align}
	\Corr(\beta_k,\beta_j|\betav_{\setminus kj}) & =-\frac{q_{kj}}{\sqrt{q_{jj}q_{kk}}} = \frac{\kappa_{kj}}{\sqrt{q_{jj}q_{kk}}} \label{par_cor}\\
	\Prec(\beta_k|\betav_{\setminus k}) & =q_{kk} /(\gamma \tau^2). \label{par_prec}
\end{align}
Thus, depending on the value of the  binary indicator  $\delta_{kj}$,  the prior   allows for high  (if $\delta_{kj}=0$) or  low (if $\delta_{kj}=1$)  positive  prior partial correlation of $\beta_k$ and $\beta_j$.


The prior  partial  precision  $\Prec(\beta_k|\betav_{\setminus k})$  can take one of $c$ different values, depending on the  binary indicators involving level $k$.  Subsuming these indicators in   $\tilde \deltav_{k}= (\delta_{k0}, \dots,\delta_{k,k-1},\delta_{k+1,k},\dots \delta_{c,k})$, the  
 minimum  value  for  $\Prec(\beta_k|\betav_{\setminus k})$ results for $\tilde \deltav_{k}= \mathbf{1}$  as  $c/(\gamma\tau^2)$, and  its maximum value, attained for $\tilde\deltav_{k}=\mathbf{0}$,  is $ r \cdot c/(\gamma\tau^2)$.  Thus the precision ratio $r$ is the ratio of maximum to  minimum prior partial  precision.
With  our choice of  $\gamma=c/2$   the partial prior precision ranges  from  $2/\tau^2$ to  $ r \cdot 2/\tau^2$, and does not depend on  the number  levels $c$.

  To illustrate the specification of the  structure matrix $\Qv(\deltav)$  we consider  a covariate with $c=3$ levels where  only one of the indicators $\delta_{kj} $, $0\le j <k \le c$ has the value $0$   whereas  all others are $1$. For a  precision ratio of $r=10000$ and $ \delta_{10}=0$   the structure matrix is given as   
$$\Qv(\deltav)= \begin{pmatrix}10002 & -1  & -1 \\
                          -1  & 3 &-1\\
                          -1 & -1 & 3  
                 \end{pmatrix}.$$ 
The marginal  prior  on $\beta_1$ is concentrated close to zero,  thus encouraging fusion to the baseline category. If  $\delta_{31}=0$,        the structure matrix is 
 $$\Qv(\deltav)= \begin{pmatrix} 10002  & -1 & -10000 \\
                                   -1 &3 & -1\\
                                   -10000 & -1& 10002
 				\end{pmatrix}.$$
Hence,  the joint prior on $(\beta_1,\beta_3)$ is concentrated close to  $\beta_1=\beta_3$ and  encourages fusion of these two effects.

The quadratic form  given in equation (\ref{eq:quadr_form}) suggests an interpretation of the  effect fusion prior in terms of conditional  Normal priors  with zero mean and precision proportional to $\kappa_{kj}$  on all  effect differences $\theta_{kj}=\beta_{k} -\beta_{j}$, $0\le j< k \le c$.  For $\delta_{kj}=0$ the prior precision is high with $\kappa_{kj}=r$  and hence  the effect difference  $\theta_{kj}$  is concentrated around zero, whereas it  is more dispersed for $\delta_{kj}=1$  where   $\kappa_{kj}=1$.   Actually, as we show in  Appendix \ref{app:pri_diff}  the effect fusion prior specified above can be derived alternatively  by first  specifying  independent spike and slab priors on  all  effect  differences $\theta_{kj}$,  and  then correcting  for  the  linear restrictions  $\theta_{kj}=\theta_{k0} -\theta_{j0}$. Thus,  the proposed prior allows for  shrinkage  of effect contrasts  to zero while taking into account their linear dependence.

As all pairwise effect contrasts  $\theta_{kj}$ are taken into account symmetrically, the effect fusion prior is invariant  to the choice of the baseline category, see  Appendix \ref{app:inv} for a formal proof. This invariance distinguishes the effect fusion prior from the conventional  spike and slab prior used for variable selection, which  allows  only  shrinkage    of
regression effects  $\beta_k=\theta_{k0}$, i.e. effect contrasts  with respect  to the baseline,  but not   all other  effect  contrasts $\theta_{kj}$, where $j>0$.

Finally, we note that from a frequentist perspective, the effect fusion  prior   can be interpreted as an adaptive quadratic  penalty with either heavy or slight  penalization of effect differences,  see equation (\ref{eq:quadr_form}).  Similarily, \cite{ger-tut:spa} in a frequentist approach  use a weighted  $L_1$ penalty on the effect differences, which has the advantage that effect differences are not only shrunken  but can actually  be set to zero.

\subsection{Prior for restricted effect fusion} \label{sec:pri_restr}
We consider now the case  where due to  available information on the structure of  levels  fusion is restricted to  specific pairs  of level effects.  A prominent example is  
an ordinal covariate where the ordering of levels suggests to allow only fusion of subsequent level effects $\beta_{k-1}$ and $\beta_{k}$.  A restriction that e.g.~$\beta_k$ and $\beta_j$ should not be fused can be implemented in our prior  in two ways:  we can   either fix the  corresponding indicator at $\delta_{kj}=1$ or  set  the corresponding element in  the prior precision matrix $\Qv(\deltav)$ to zero. Whereas $q_{kj}=0$  is a hard restriction which implies conditional independence of   $\beta_k$ and $\beta_j$, setting $\delta_{kj}=1$ is  a soft restriction which implies that effects $\beta_k$ and $\beta_j$ are still  smoothed to each other. 

The implementation of soft restrictions is straightforward, but  (hard) conditional independence restrictions require  slight modifications  in  the definition of the  structure matrix $\Qv(\deltav)$, the vector of indicators $\deltav$ and the constant  $\gamma$. To specify hard fusion restrictions we
  introduce  a  vector $\zetav$  of  indicators $\zeta_{kj}$,  which are  defined  for each effect difference $\theta_{kj}$. The elements of  $\zetav$ are fixed and indicate  whether an effect difference is subject to fusion (for  $\zeta_{kj}=1$) or not (for   $\zeta_{kj}=0$). Deviating from unrestricted effect fusion considered in Section \ref{sec:pri_nom}, we define a stochastic indicator $\delta_{kj}$ only for those effect differences where $\zeta_{kj}=1$ and hence the dimension of $\deltav$ is $d=\sum_{k=1}^c \sum_{0\le j< k}\zeta_{kj}$.
 
To allow off-diagonal elements of  the prior precision to be zero, we set
    $$ q_{kj}  =\begin{cases}  -\kappa_{kj} \quad & \text{if }  \zeta_{kj } =1\\
    \phantom{-} 0 & \text{if  }  \zeta_{kj} =0\end{cases}$$
and  $q_{jk}=q_{kj}$. Thus, $q_{kj}$ takes the value zero  if $\zeta_{kj}=0$ and  $ -\kappa_{kj}$ otherwise. Accordingly, the diagonal elements are specified as      
$$ q_{kk} = \begin{cases}  \kappa_{k0}   -\sum_{k \neq j} q_{kj}   \quad &   \text{if }  \zeta_{k0 } =1\\
   \phantom{ \kappa_{k0}} -\sum_{k \neq j} q_{kj}  & \text{if }  \zeta_{k0 } =0.\end{cases} $$
 
As noted above, an important special case is an ordinal covariate,  where  fusion of effects can be   restricted to adjacent categories, see \cite{ger-tut:pen}. We define 
$$\zeta_{kj}= 
\begin{cases} 
	1 \qquad & \text{for } j=k-1\\
   	0 & \text{otherwise}.
\end{cases}$$ 
Thus,  the vector of indicators $\deltav$  has only $d=c$ elements, $\deltav=(\delta_{10}, \dots \delta_{c,c-1})$ and $\Qv(\zetav,\deltav)$  is a tri-diagonal matrix with elements
\begin{equation} \label{eq:priQord}
	\Qv(\zetav,\deltav)=
	\begin{pmatrix} \kappa_{10}+\kappa_{21}    &-\kappa_{12}  &  \dots& 0 & 0\\
	-\kappa_{21} &  \kappa_{21}+\kappa_{32}     & \dots & .& 0\\
	0 &  -\kappa_{32}& &\dots& .   \\
	\vdots & \vdots & \vdots   & \ddots & \vdots \\
	  . & . & \hdots & \kappa_{c-1,c-2}+\kappa_{c,c-1} & -\kappa_{c,c-1}\\
	0 &0  &\hdots & -\kappa_{c,c-1} & 
	\kappa_{c,c-1} 
	\end{pmatrix}.
\end{equation}
In this case, the  maximum value of a diagonal element is  $q_{kk}=2r$   and therefore we  set $\gamma=1$.

It is easy to show that this specification of $\Qv(\zetav,\deltav)$ corresponds to a random walk prior   on the regression effects:
$$ \beta_k =\beta_{k-1} + \theta_k, \qquad \theta_k \sim \Normal{0,\tau^2/ \kappa_{k,k-1}},$$
with initial value $\beta_0=0$.
Due to the spike and slab structure, this prior allows for adaptive smoothing, with almost no  smoothing for $\delta_{k,k-1}=1$ and pronounced smoothing for $\delta_{k,k-1}=0$. 

Another special case of a restricted effect fusion prior is the standard spike and slab prior used for   variable selection, which encourages only fusion to the baseline,   i.e.~shrinkage  of $\beta_{k}$  to $\beta_0=0$. In our framework the spike and slab prior  is recovered  with   
$$\zeta_{kj}=\begin{cases} 1 \qquad  & \text{for }  j=0\\
   0 & \text{otherwise}\end{cases}$$ 
and  $\gamma=1$. Therefore the off-diagonal elements of $\Qv(\zetav,\deltav)$ are zero and
$q_{kk}=  \kappa_{k0}$.

\subsection{Prior on the indicator variables}  \label{sec:pri_ind}

In variable selection    elements of $\deltav$  are usually assumed to be conditionally independent  a priori with $ p(\delta_{kj} =1)  =\omega$,  where $\omega$ is  either fixed or assigned a hyperprior $\omega \sim \mathcal{B}(v_{0}, w_{0})$.   This would  be possible also for the effect fusion prior, but from a computational point of view  a more convenient choice  is to set
\begin{equation} \label{eq:pridel} p(\deltav)\propto |\Qv(\deltav)|^{-1/2}r^{\sum(1-\delta_{kj})/2}. \end{equation}
Thus,   the determinant  of $ \Qv(\deltav)$ cancels  out in the joint prior  of regression effects and indicators $p(\betav, \deltav|\tau^2)$, which results as
 \begin{equation}\label{eq:postdel} p(\betav,\deltav| \tau^2 ) = p(\betav|\deltav,\tau^2)p(\deltav)= (\frac{1}{\gamma \tau^2})^{c/2 }\exp\big(-\frac{\betav' \Qv(\deltav) \betav}{2\gamma \tau^2}  \big) (\sqrt{r})^{\sum(1-\delta_{kj})}.
 \end{equation} 
 This prior has attractive features: Firstly, as 
 $p(\betav, \deltav|\tau^2)$ can be factorized as 
\begin{align}\label{eq:postdel1} p(\betav, \deltav|\tau^2) & \propto \prod_{k\neq j}  (\sqrt{r})^{1-\delta_{kj}}  \exp\Big(-\frac{(\beta_k-\beta_j)^2}  {2\tau^2 \gamma} \big(\delta_{kj}+r(1-\delta_{kj}) \big) \Big) \propto \\ & \propto \prod_{k\neq j} p(\theta_{kj}, \delta_{kj}|\tau^2),\label{eq:postdel1a}
 \end{align} 
 the  effect differences $\theta_{kj}$ and indicators $\delta_{kj}$ are jointly independent across  all pairs $0\le j < k \le c$  conditional on the scale parameter $\tau^2$. Secondly,   conditioning on the corresponding effect  difference $\theta_{kj}$
 the  conditional  prior of the indicator $\delta_{kj}$, 
  \begin{equation} \label{eq:condpri}p(\delta_{kj}=1|\betav,\tau^2)=\frac{p\big(\theta_{kj}| \Normal{0,\gamma \tau^2}\big)}{p\big(\theta_{kj}| \Normal{0,\gamma \tau^2}\big)+ p\big(\theta_{kj}| \Normal{0,\gamma \tau^2/r}\big)}, \end{equation}  is identical  for all  pairs of indices $k,j$ subject to fusion.

 The properties of the prior $p(\deltav)$ given in equation (\ref{eq:pridel}) depend on the concrete specification of $\deltav$.  For the special cases of restricted effect fusion discussed above,   fusion of subsequent levels of an ordinal  covariate  with  $\deltav=(\delta_{10}, \dots \delta_{c,c-1})$  and  $\Qv(\deltav)$ specified in equation (\ref{eq:priQord}),  and  variable selection
with $\deltav=(\delta_{10}, \dots \delta_{c,0})$ and  $\Qv(\deltav)=\diag(\kappa_{10},\dots, \kappa_{c0})$, the prior
  $p(\deltav)$
 is uniform over all $2^c$ different values of 
 $\deltav$,
 $$p(\deltav) \propto 1,$$
 see Appendix \ref{app:pridel_ord}  for a formal proof.  
 
 In contrast, for a nominal covariate with unrestricted effect fusion   $p(\deltav)$ specified   in equation  (\ref{eq:pridel})   favours sparse 
 models. As  $\Qv(\deltav)$ has a more complicated structure  in this case (see formula  (\ref{eq:Qmat}) in the Appendix), its determinant  is not available in closed form for arbitrary $\deltav$, but  we can compare the  full model, where  $\deltav=\mathbf{1}$ and the null model where $\deltav=\zerov$ and hence  all effects are sampled from the spike component.
   As $|\Qv(\zerov)|=| r \Qv(\mathbf{1})|$,  we get
    $$\frac{p(\deltav=\zerov)}{p(\deltav=\mathbf{1})}= \frac{|\Qv(\mathbf{1})|^{1/2} }{|\Qv(\zerov)|^{1/2}} \cdot r^{d/2}=  r^{-c/2} r^{d/2}= r^{c(c-1)/2}.$$ 
Thus, a priori the null model is increasingly favoured over the full model with higher number of  levels and higher precision ratio.

\begin{figure}
 \centering
	\includegraphics[height=5cm, width=5cm]{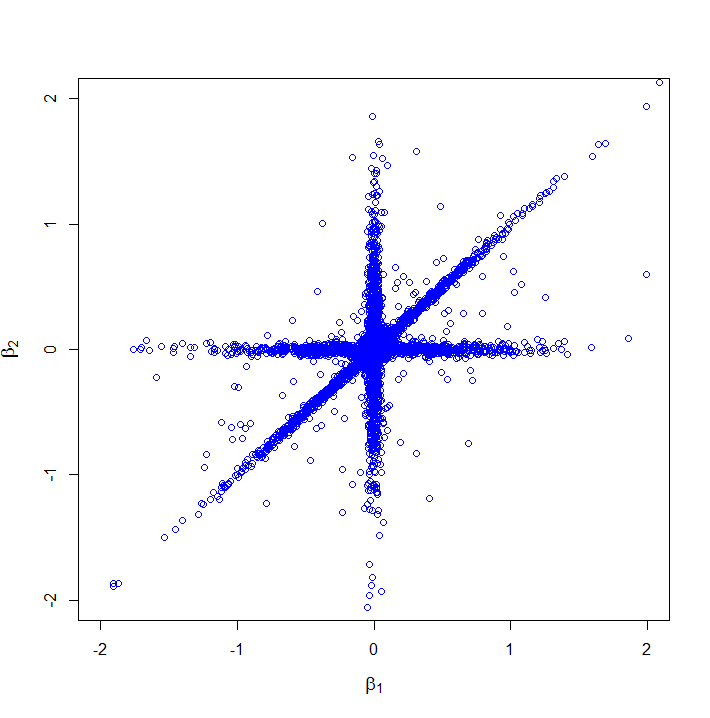}
	\includegraphics[height=5cm, width=5cm]{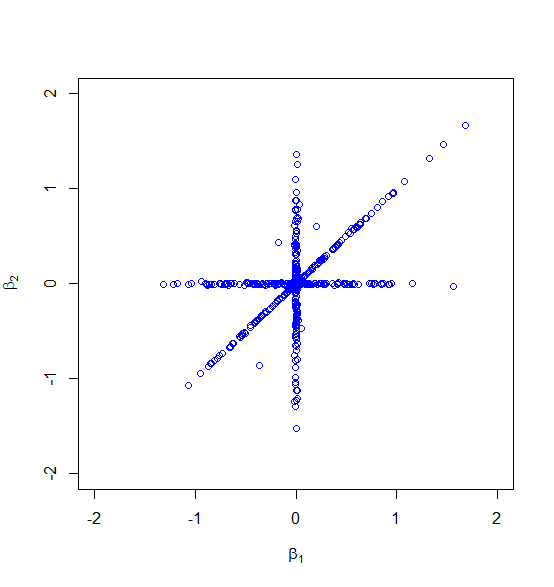} \\	
	\caption{\label{fig:prisim} Simulation  from the effect fusion prior: Plot of  $(\beta_1,\beta_2)$ for $c=3$ and  hyperparameters $g_0=5$, $G_0=2$,  $r=1000$ (left) and $r=10000$ (right)  }					
\end{figure}

 Figure \ref{fig:prisim}  shows simulations from the marginal prior $ p(\betav)$ for a categorical regressor with $c=3$ levels (except baseline) for $r=1000$ and $r=10000$. Due to the  symmetry of the prior with respect to all level effects   only plots for $(\beta_1,\beta_2)$ are presented.  
  The prior is concentrated at configurations where  all or a subset of the regression effects is zero,  or where two effects are equal, $\beta_j=\beta_k$ and concentration at these configurations increases with the precision ratio $r$.

\begin{figure}
\centering
	\includegraphics[height=5cm, width=5cm]{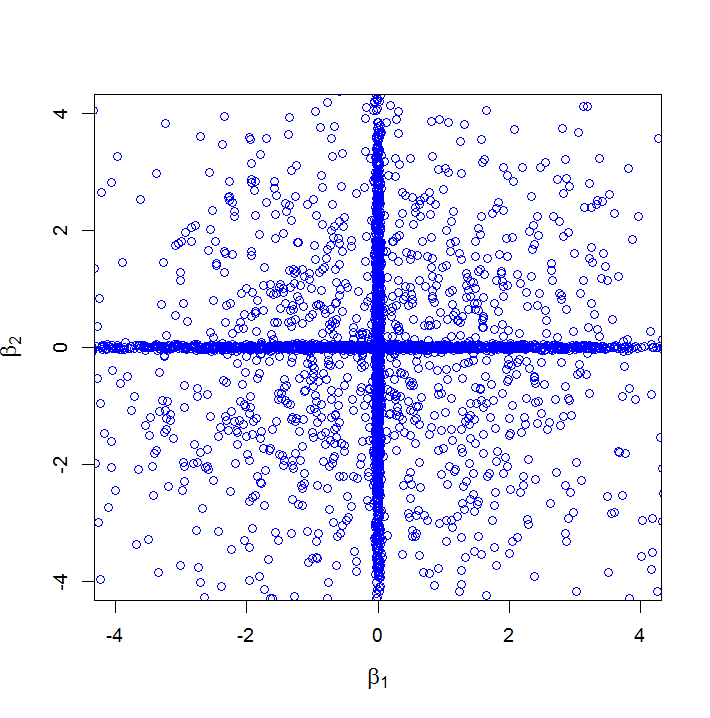}
	\includegraphics[height=5cm, width=5cm]{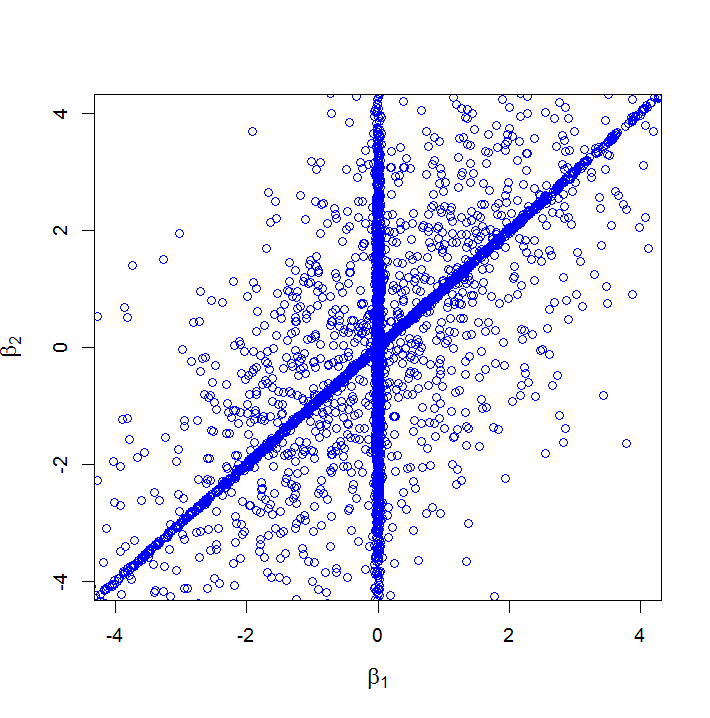} \\
	\caption{\label{fig:prisim_restr} Simulation from the prior: Plot of $(\beta_1,\beta_2)$ for $c=3$ and hyperparameters  $g_0=5$, $G_0=20$ and $r=10000$. Left: variable selection prior, right: ordinal fusion prior   }					
\end{figure}

Corresponding plots for simulations from the variable selection prior, where fusion is restricted to 
$\beta_k=0$, and the ordinal  fusion prior with fusion restricted to $\beta_k=\beta_{k-1}$
are shown in Figure \ref{fig:prisim_restr}. Compared to unrestricted effect fusion,  sparsity is much less pronounced in  these two priors.
 

 \subsection{Choice of hyperparameters}\label{sec:hyp}
To provide a rationale for the choice of the  hyperparameters $r, g_0$ and $G_0$  we focus on the joint  marginal prior distribution of one  effect difference  and the corresponding indicator  $0\le j < k \le c$. As already noted above,  the random variables  $(\theta_{kj},\delta_{kj})$ are independent and  identically distributed for all pairs of indices $k,j$   conditional on the scale parameter $\tau^2$.   Marginalized over $\delta_{kj}$ and  $\tau^2$ the prior  an effect difference $\theta_{kj}$ is  a spike and slab distribution,  where both the spike and the slab are scaled t-distributions with $2g_0$ degrees of freedom and scale parameter  $\sigma=\sqrt{G_0/g_0}$ (for $\delta_{kj}=1)$  and  $\sigma/\sqrt{r}$ (for $\delta_{kj}=0)$, respectively. 
For effect fusion we follow the standard choice in variable selection to set $g_0=5$ \citep{fah-etal:bay, sch-etal:spi} where
tails of  spike and slab  fat enough to avoid MCMC mixing problems when the indicator $\delta_{kj}$ is  sampled conditional on  $\theta_{kj}$.
 
 Our modelling goal is to  allow for  fusion of  level effects with negligible difference  while level effects with relevant difference
 should be modelled seperately. To avoid misclassification of relevant differences as negligible, which we call \emph{false negatives}  or of negligible effects as relevant, i.e.  \emph{false positives},  the scale parameter $G_0$ and the precision ration $r$ could be chosen by specifiying the   conditional fusion probability   
  \begin{equation*} P(\delta_{kj}=0| \theta_{kj})=\frac{1}{1+\frac{t_{2 g_0}(\frac{\theta_{kj}}{\sigma})}{\sqrt{r}t_{2 g_0}(\sqrt{r}\frac{\theta_{kj}}{\sigma})}}\end{equation*} 
 for two values of $\theta_{kj}.$
 
\begin{figure}
	\includegraphics[height=5cm, width=6cm]{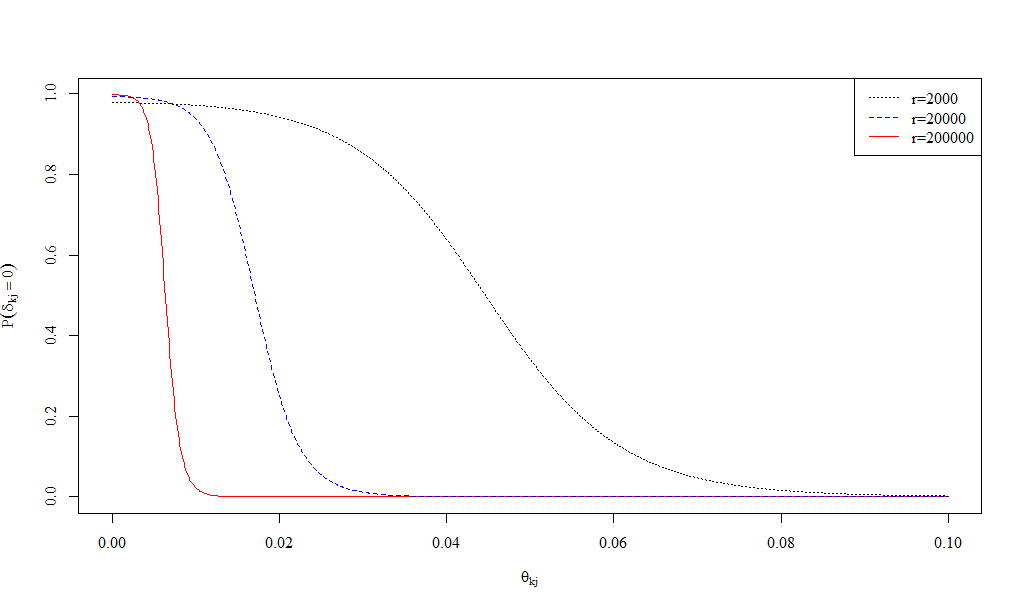}
	\includegraphics[height=5cm, width=6cm]{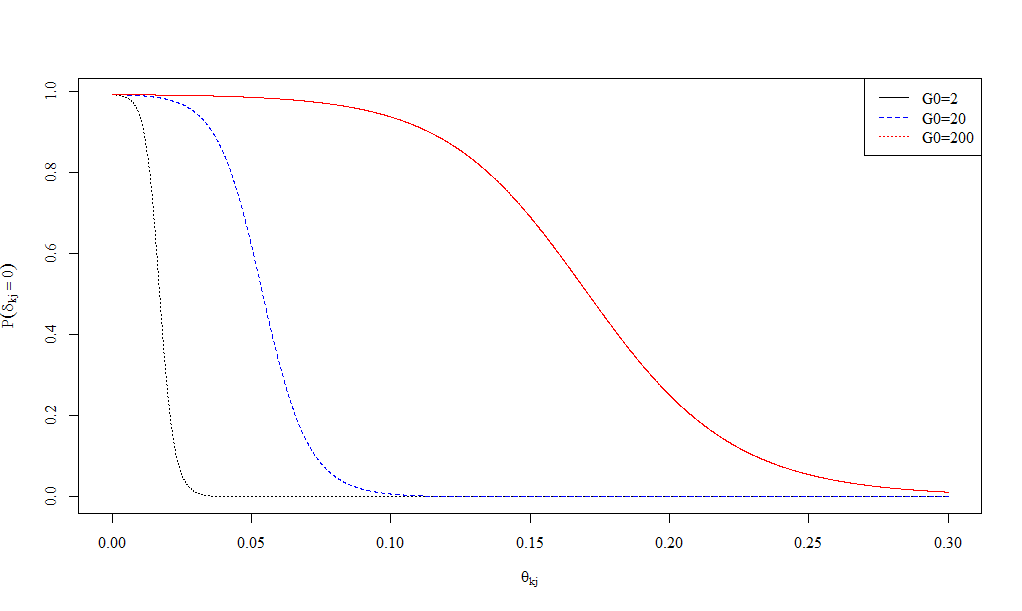}
		\caption{\label{fig:prifuse} Fusion probabilities for  $g_0=5$. Left:  $G_0=2$ and different values of $r$, right: $r=20000$ and different values of $G_0$}					
\end{figure}

   Figure \ref{fig:prifuse} shows plots of  the fusion probabilities as a function of the effect difference  for various values of $r$ and $G_0$ with $g_0=5$.  For fixed $\theta_{kj}$ the fusion probability  $P(\delta_{kj}=0|\theta_{kj})$ decreases with $r$ and increases with $G_0$ which suggests  to choose a large value for $r$ large and a small value for $G_0$.      
    However  for smaller values of shrinkage effect differences to zero is more pronounced even under the slab, which might hamper detection of small effect differences and hence $G_0$ has to be chosen carefully to represent the scale of relevant effects. 
   
  We will investigate  different values for both hyperparameters $G_0$ and $r$ in  the simulation study in Section \ref{sec:sim_hyp}. 
 As suggested by   a referee,  we investigate also a hyperprior on $G_0$, where  a convenient choice is an exponential prior $G_0 \sim \Exp{\lambda}$ with mean $\lambda$.  We tried also to put a hyperprior on $r$,  which did however not prove useful as the additional  flexibility resulted in all effect differences being assigned to the spike component.

\section{Posterior inference} \label{sec:inference}
 Our goal  is   posterior inference  for the parameters of 
  linear regression model
  $$\yv=\Xv \betav+ \errorv, \quad \errorv \sim \Normal{0,\sigma^2 \Iv}$$
    with  prior 
    $$p(\betav,\deltav, \tauv^2, \sigma^2) =  p(\betav|\deltav,\tauv^2)p(\deltav)p(\tauv^2)p(\sigma^2).$$
 Here  the regressor matrix is  $\Xv=[\mathbf{1}, \Xv_1,\dots, \Xv_p]$ and the vector of regression effects is $\betav=(\mu,\betav_1',\dots, \betav_p')'$. The vector
 $\deltav=(\deltav_1',\dots, \deltav_p')'$   stacks the vectors of binary indicators $\deltav_h$, $h=1,\dots,p$ and   $\tauv^2=(\tau_1^2,\dots, \tau_p^2)'$ subsumes the scale parameters $\tau_h^2$. 
  The prior for $\betav$ is   multivariate Normal, $\betav|\deltav,\tauv^2 \sim \Normal{\zerov, \Bv_0(\deltav,\tauv^2)}$ with   covariance matrix $\Bv_0(\deltav,\tauv^2)$, which is  block-diagonal with elements $M_0$ (for the intercept) and $\Bv_{h0}(\deltav_h,\tau_h^2)$, $h=1,\dots,p$.
  
The resulting posterior distribution is proper even for the  partial improper prior with $p(\mu, \sigma^2)\propto  \frac{1}{\sigma^2}$ under conditions given by  a theorem of \cite{sun-etal:prop},   see  Appendix \ref{app:prop} for details.

Posterior inference can be accomplished by sampling  from  the posterior distribution using MCMC methods.
For the  prior distributions specified above posterior inference  for the model parameters is feasible by a Gibbs sampler, where  the full conditonals  are  standard distributions. The  sampling scheme is outlined in Section \ref{sec:MCMC}.
Model averaged estimates of the parameters can be obtained as the means of the MCMC draws but  often the goal is to select a final model with eventually fused levels. In Section 4.2. we discuss  how this model selection problem can be addressed in   Bayesian decision theoretic approach by choosing an appropriate loss function. 

\subsection{MCMC scheme} \label{sec:MCMC}
We initialise  MCMC by choosing starting values   the error variance $\sigma^2$,  the indicators $\deltav$ and the scale parameters $\tauv^2$ and compute  the prior covariance  matrix $\Bv_0(\deltav,\tauv^2)$.

MCMC  then iterates the following steps:
\begin{enumerate}
\itemsep 0pt		
	\item[(1)]  Sample the regression coefficients $\betav$ from the  full conditional Normal posterior $p(\betav|\sigma^2, \deltav,\tauv^2,\yv)$. 
				
	\item[(2)] Sample  the error variance $\sigma^2$ from  the  full  conditional Inverse 
	Gamma distribution  $p(\sigma^2|\betav,\deltav,\tauv^2, \yv)$.
	
	\item[(3)] For $h=1,...,p$: Sample the scale parameter $\tau^2_h$ from 
	the full conditional  Inverse Gamma distribution $p(\tau^2_h|\betav,\sigma^2,\deltav_h)$.

     \item[(4)]  For all $h \in \{1,\dots , p\}$ with   a hyperprior specified on $G_h$:  Sample $G_h$ from $p(G_h|\lambda_h,\tau_h^2)$
		
	\item[(5)]   For  $h=1,\dots,p$: Sample $\deltav_{h}$  independently 
  from its full conditional posterior $p(\deltav_h|\betav_h,\tau^2_h,\sigma^2,\yv)$.

  \item[(6)]  Update the prior covariance matrix $\Bv_0(\deltav, \tauv^2)$.        
\end{enumerate}

  Due to the hierarchical structure of the prior,  the posterior of $\deltav_h$  in sampling step (5)  depends only on $\betav_h$ and $\tau^2_h$  and as discussed in Section
  \ref{sec:pri_ind}  elements of   $\deltav_h$ are conditionally independent given  $\tau^2_h$ and $\betav_h$. Therefore     all binary  indicators $\delta_{h,kj}$ can  be sampled independently  from $p(\delta_{h,kj}=1|\theta_{kj}=\beta_{h,k}-\beta_{h,j},\tau_h^2) $, which is  given in equation  (\ref{eq:condpri}).

Full details of the sampling steps are provided in Appendix \ref{app:detinf}. The sampling scheme is implemented in the {\tt R} package {\tt effectFusion} \citep{pau-etal:eff}.

Compared to  posterior inference with a  standard Normal prior on the regression effects $\betav$,   only  steps (3) - (6)  have to be added under the 
effect fusion prior. These steps are fast for nominal covariates  that are typical in  applications with $2-50$ levels. However, as all pairwise differences are assessed in each sweep of the sampler computation times become prohibitive for covariates with 100 or more levels. For a data set of $n=10000$ observations   and one nominal covariate 1000 MCMC iterations take  1.5 seconds  for a covariate with $c=20$ levels, but  roughly 25 min.  for $c=100$ levels on a standard laptop  with Intel i7-5600U processor with 2.60 GHz and 16 GB RAM. More details on computation times are given in  Appendix \ref{app:comptime}.          

 As already noted above, due to the hierarchical structure of the effect fusion prior  the full conditionals of the parameters $\deltav_h,\tau_h^2$ and $G_h$  depend only on the  regression effects $\betav_h$  and not on the  data likelihood. Thus, implementation of effect fusion for categorical covariates  is straightforward in any type of regression model with an  additively structured linear predictor.

\subsection{Model selection} \label{sec:model_sel}
If the goal is to select a final model, e.g. to be used for prediction,   a Bayesian decision theoretic approach  requires to  choose an appropriate loss function. A particularly  appealing loss function for effect fusion is a special case Binder's loss \citep{bin:bay} which is  used in \cite{lau-gre:bay} for Bayesian model based clustering of observations. It  considers pairs of items and penalizes incorrect clustering, which occurs when  two items which should not be clustered are assigned to the same cluster or when two items which should be clustered are assigned to different clusters. For effect fusion incorrect clustering corresponds to classifying an effect difference   falsely as negative  or falsely as positive, respectively.
 
 Binder's loss is given as
$$\mathcal{L}(\zv, \zv^*)= \sum_{j \neq k } \big( \ell_1 \mathbf{I}_{\{z_{k}=z_{j}\}}\mathbf{I}_{\{z^*_{k} \neq z^*_{j}\}} +\ell_2  \mathbf{I}_{\{z_{k}\neq z_{j}\} }\mathbf{I}_{\{z^*_{k} = z^*_{j}\}} )  $$
where   $\zv$ denotes the true  and $\zv^*$ the proposed  clustering   and
 the constants $\ell_1$  and $\ell_2$
are  misclassification costs. For $\ell_1=\ell_2$ the expected posterior loss  results as 
\begin{equation} \E(\mathcal{L}(\zv, \zv^*)|\yv)=\sum_{j \neq k} | \mathbf{I}_{\{z^*_{k} = z^*_{j}\}}-\pi_{kj}|  \label{eq:exp_binder}\end{equation}
where  $\pi_{kj}=P(z_{k} = z_{j}|\yv)$ denotes  the $(k,j)$ element of the posterior similarity matrix, see  \cite{fri-ick:imp}.
The Bayes optimal action, i.e.  the clustering  which minimizes  the expectation of Binder's loss, can be determined by minimizing
\begin{equation} \sum_{j \neq k}  \mathbf{I}_{\{z^*_{k} = z^*_{j}\}}(\frac{1}{2} - \pi_{kj}).\label{eq:minbinder} \end{equation}
For this minimization problem \cite{lau-gre:bay} use  an algorithm based on integer programming, which is implemented in the function 
{\tt minbinder} ({\tt R} package {\tt mcclust}).
 
 We determine the optimal fusion model  with respect to the expected posterior Binder loss (\ref{eq:exp_binder}) for each covariate  $C_h$ seperately and  approximate the elements $\pi_{h,kj}$ of the corresponding posterior similarity matrix  using  $M$ MCMC draws (after burnin) by 
 $$\hat{\pi}_{h,kj} = \frac{1}{M} \sum_{m=1}^M \delta_{h,kj}^{(m)}.$$

 Finally,  we refit the  selected model   with dummy-coded regression coefficients for the fused levels
under a flat Normal prior $\Normal{0, \Iv B_0}$.

\section{Simulation study} \label{sec:simulation}

To  investigate the performance of the proposed method  we conducted a simulation study  with a  similar set-up as in \cite{ger-tut:spa} where we 
compare results of the finally selected model under  the effect fusion prior with respect to parameter estimation, correct effect fusion and predictive performance  to various other approaches: penalized regression  (\emph{Penalty}), the Bayesian lasso (\emph{BLasso}), the Bayesian elastic net (\emph{BEN}), the group lasso (\emph{GLasso}),  the sparse group lasso (\emph{SGL}) and the Bayesian Sparse Group Lasso (\emph{BSGS}).
 Additionally, we include Bayesian regularization via graph Laplacian (\emph{GLap}), proposed in  \cite{liu-etal:bay}, where 
the prior is also specified directly on  the elements of the prior  precision matrix, however with the 
 goal to identify conditional independence by shrinking off-diagonal elements to zero. A list of the papers introducing these methods and  the related {\tt R} packages is given in the Appendix \ref{app:sim_meth}. 
For comparison  we also  fit  the full model (\emph{Full}) with separate dummy variables for each level and the true model (\emph{True}), i.e.~the model  where correct fusion is assumed to be known.

\subsection{Simulation set-up}

We   generated 100 data sets with $n=500$ observations from the Gaussian linear regression model  (\ref{model1}) with intercept $\mu=1$,  a standard Normal error $\varepsilon \sim \Normal{0,1}$ and fixed design matrix $\Xv$.  We use four ordinal  and four nominal predictors, where two regressors have eight  and two have four categories
for each type of covariate (ordinal and nominal). Regression effects are set to  $\betav_1 = (0,1,1,2,2,4,4)$ and $\betav_3 = (0,-2,-2)$  for the ordinal,  to $\betav_5 = (0,1,1,1,1,-2,-2)$ and $\betav_7 = (0,2,2)$ for the nominal covariates,  and $\betav_h=\zerov$ for $h=2,4,6,8$. Levels of the predictors are generated with probabilities $(0.1,0.1,0.2,0.05,0.2,0.1,0.2,0.05)$   and  $(0.1,0.4,0.2,0.3)$  for regressors with eight and four levels, respectively.

To perform effect fusion, we specify a Normal prior with variance  $B_0=10000$ on the intercept and the improper prior $p(\sigma^2) \propto 1/\sigma^2$ (which corresponds to an  Inverse Gamma distribution  with parameters $s_0=S_0=0$) on  the  error variance $\sigma^2$.  For each covariate $h$, the hyperparameters are set to $ G_{h0}=20$ and $r=20000$,  but we investigate also various other
 values for both parameters and an exponential hyperprior on $G_{h0}$ with $\lambda=\E(G_{h0})=2$ in Section \ref{sec:sim_hyp}. 

 MCMC  is run for 10000  iterations after burnin of 5000 to perform model selection for each data set.  Models \emph{Full} and \emph{True} and the refit of the selected model are estimated under a flat Normal prior $\Normal{0,\Iv B_0}$  on the regression coefficients where $B_0=10000$ with   MCMC  run  for 3000 iterations (after a burnin of 1000). The tuning parameters of the frequentist methods \emph{Penalty} and  \emph{GLasso}  are selected automatically via cross-validation in the corresponding {\tt R} packages. For \emph{SGL}  we  choose  the penalty parameter via  cross-validation  in the range  from $0.00005$ to $0.05$.  
 For the Bayesian methods, we use the default prior parameter settings in the code (for \emph{GLap}) and the {\tt R} packages {\tt monomvn} and {\tt EBglmNet}  and estimate  the regression coefficients by  the posterior means.  For \emph{BSGS} which is tailored to sparse group selection of numeric regressors  in a model with no intercept  the recommendation in \cite{lee-chen:bay} to demean the response is not useful in our setting and hence we set the prior inclusion probability for the intercept to 0.99 (which due to implementation specifics resulted in a higher posterior inclusion probability than a value of 1)  and used the default value of 0.5 for all other covariates.

\subsection{Simulation results}
\begin{figure}
	\includegraphics[scale=0.26]{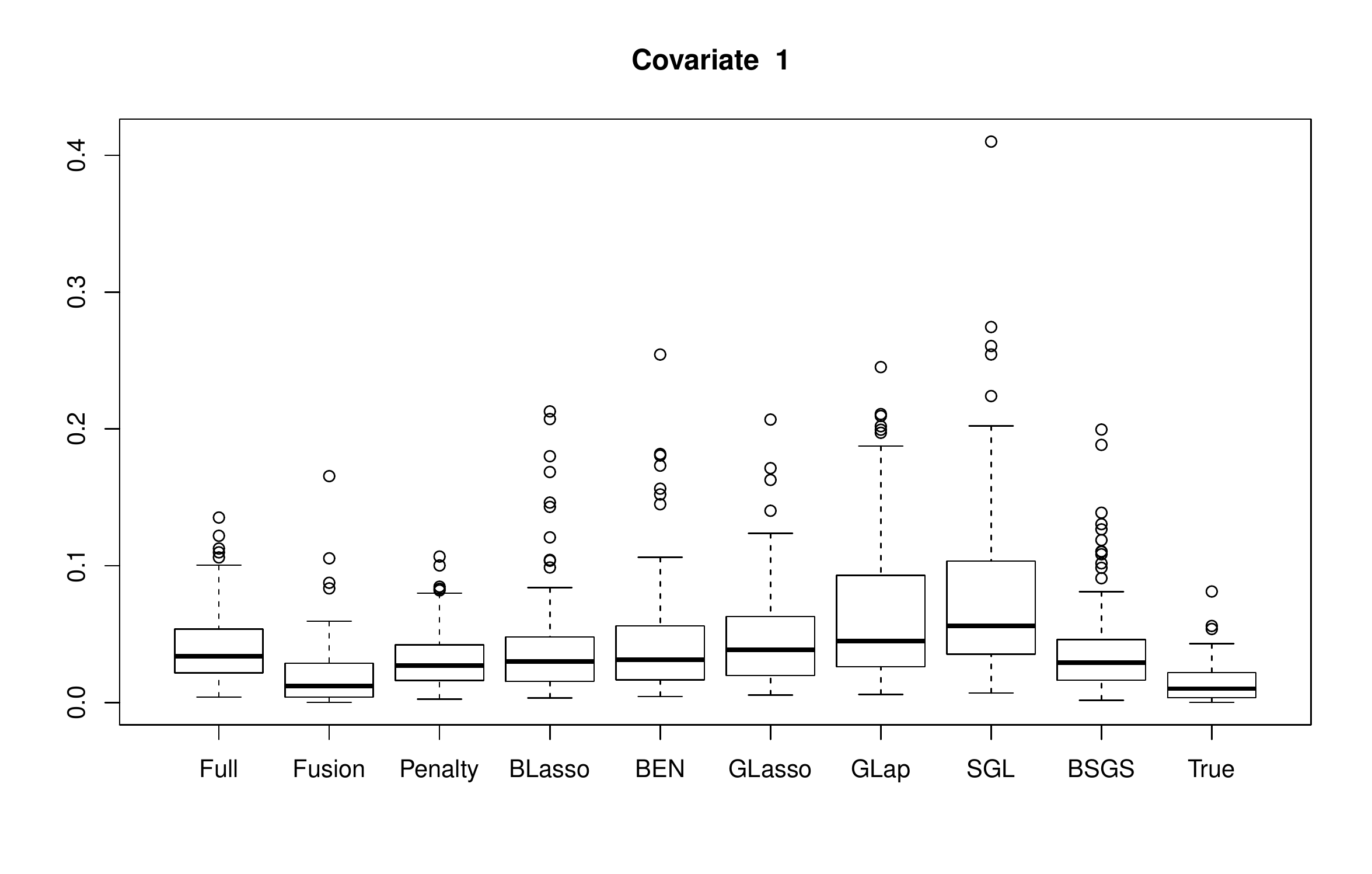}
	\includegraphics[scale=0.26]{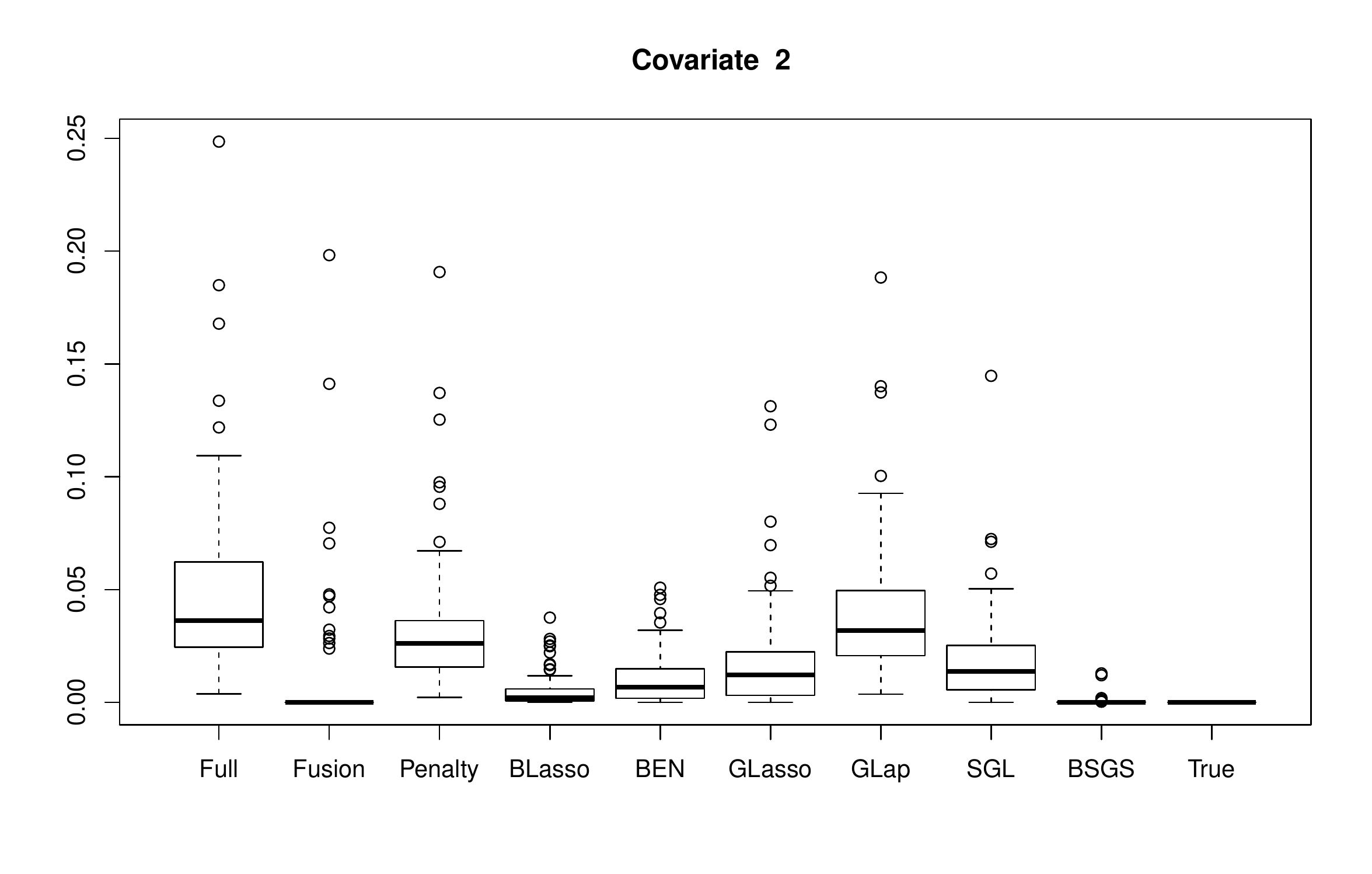} \\
	\includegraphics[scale=0.26]{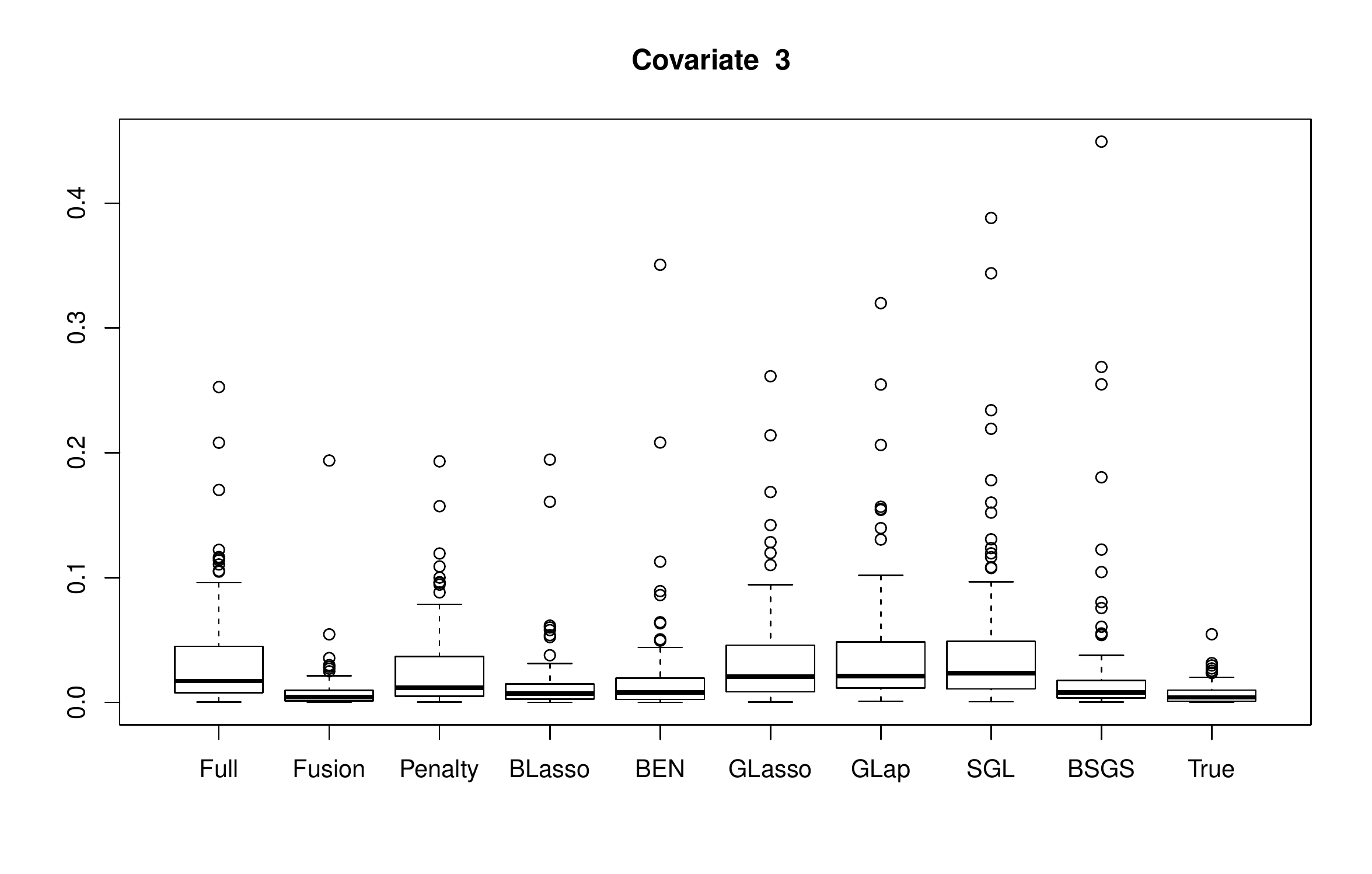}
	\includegraphics[scale=0.26]{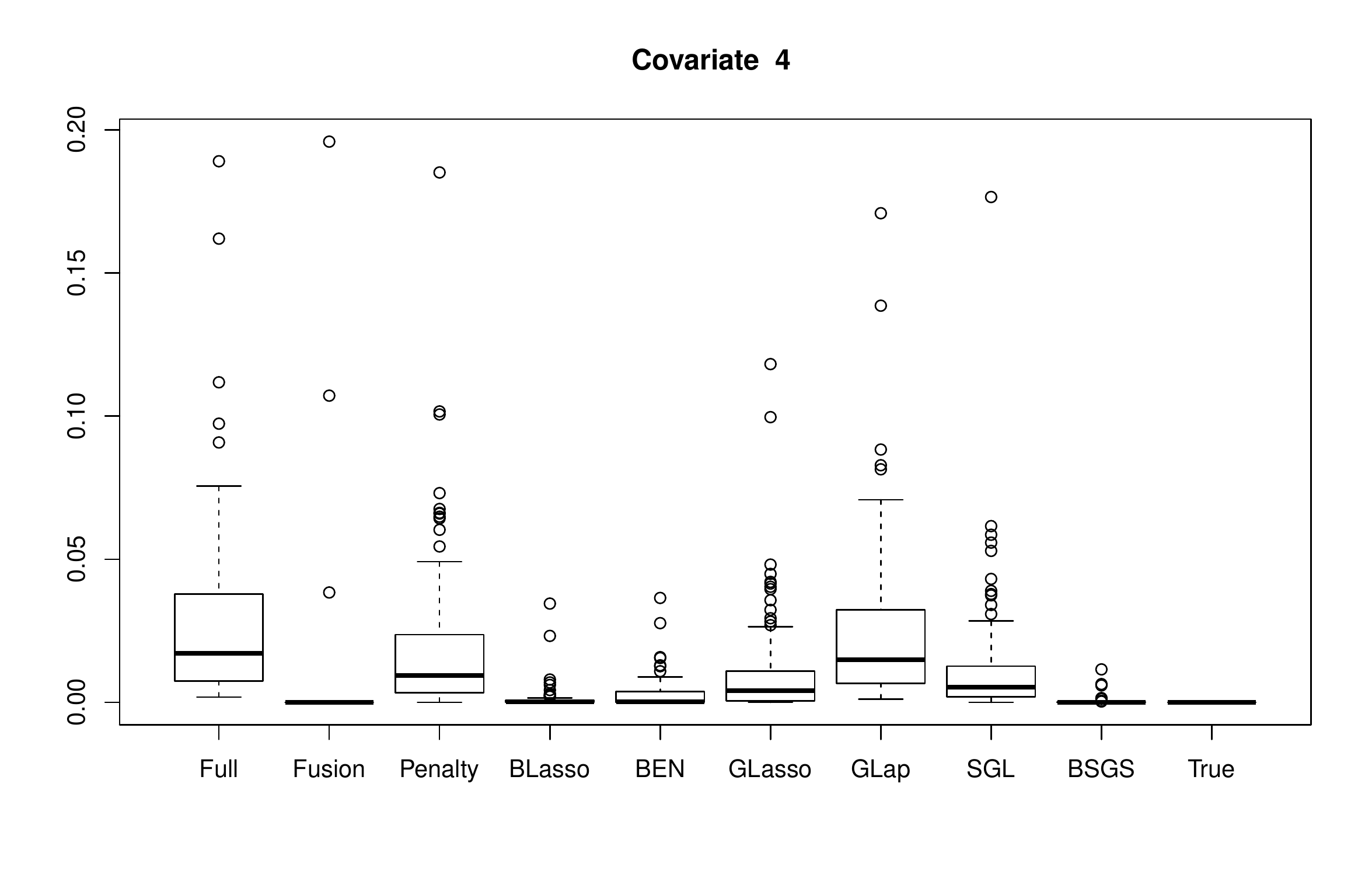} \\
	\includegraphics[scale=0.26]{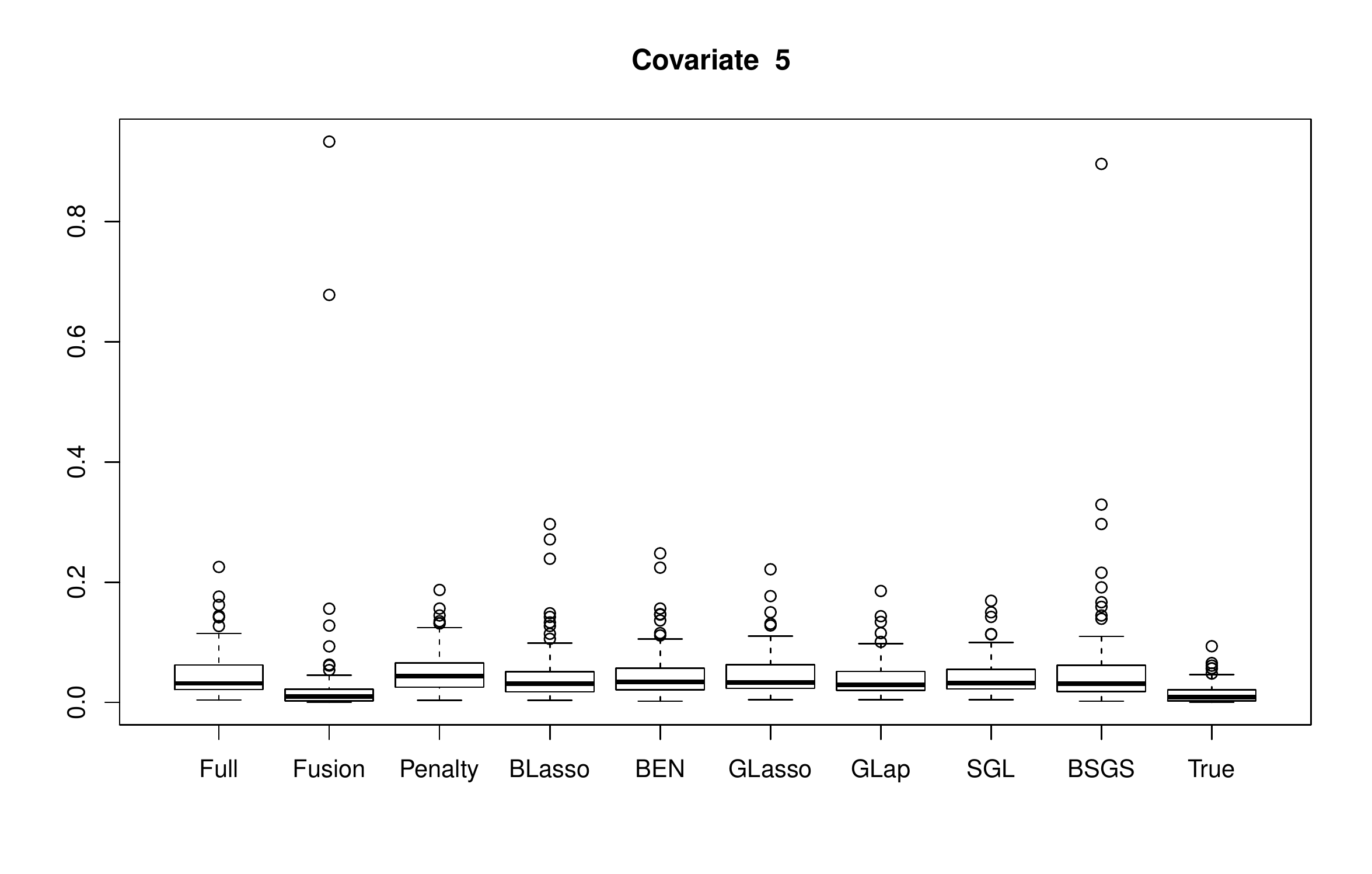}
	\includegraphics[scale=0.26]{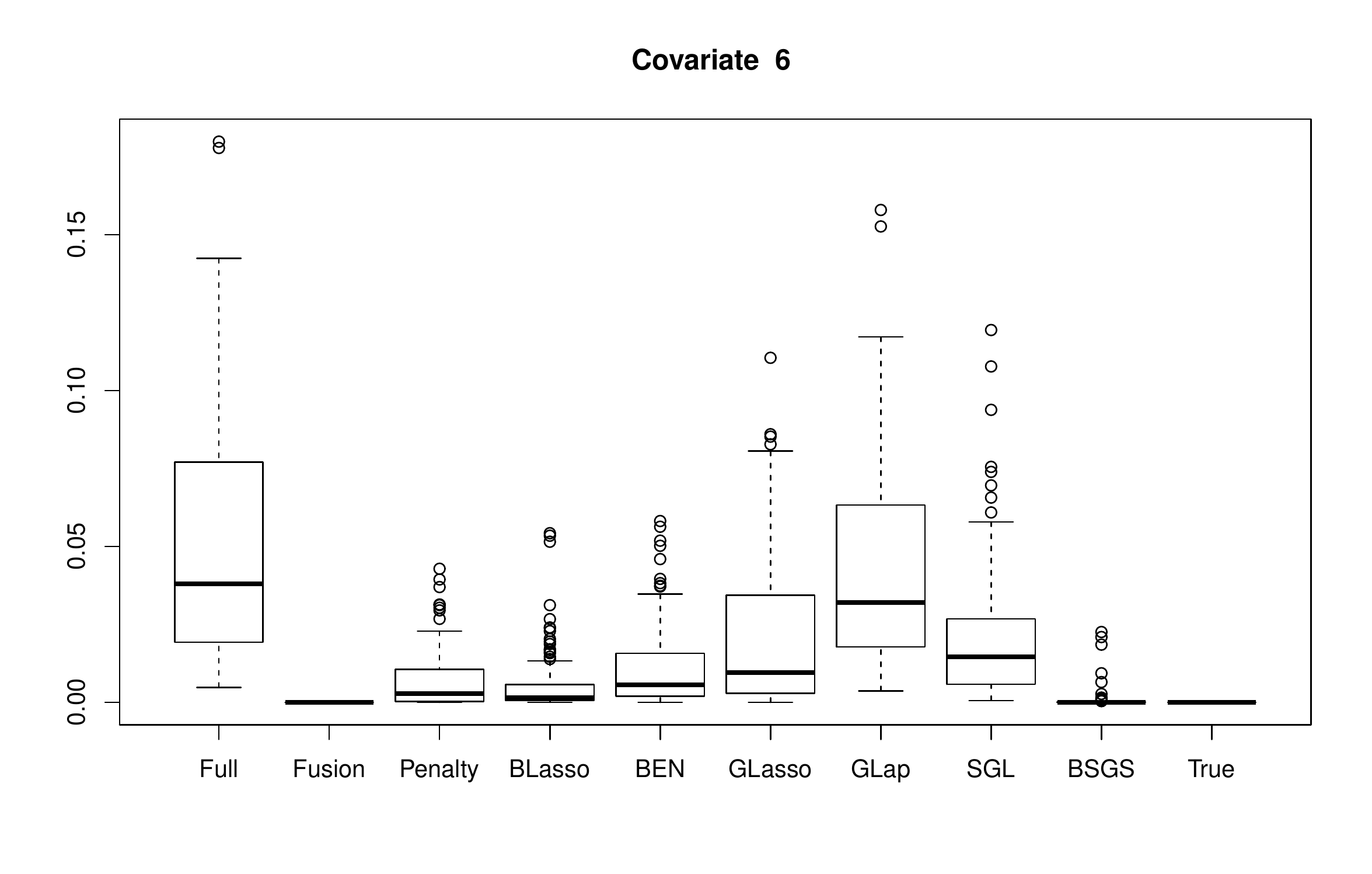} \\
	\includegraphics[scale=0.26]{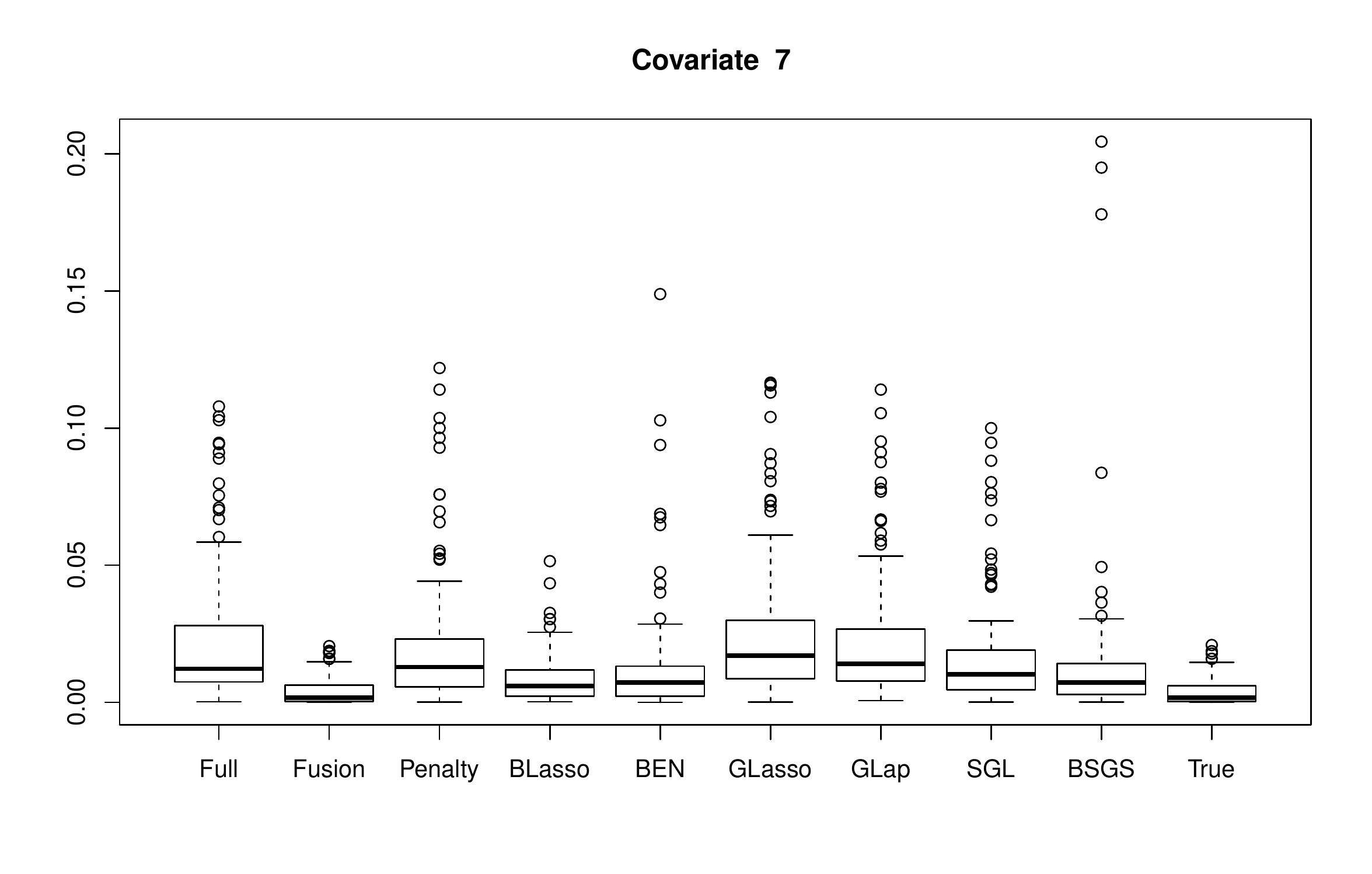}
	\includegraphics[scale=0.26]{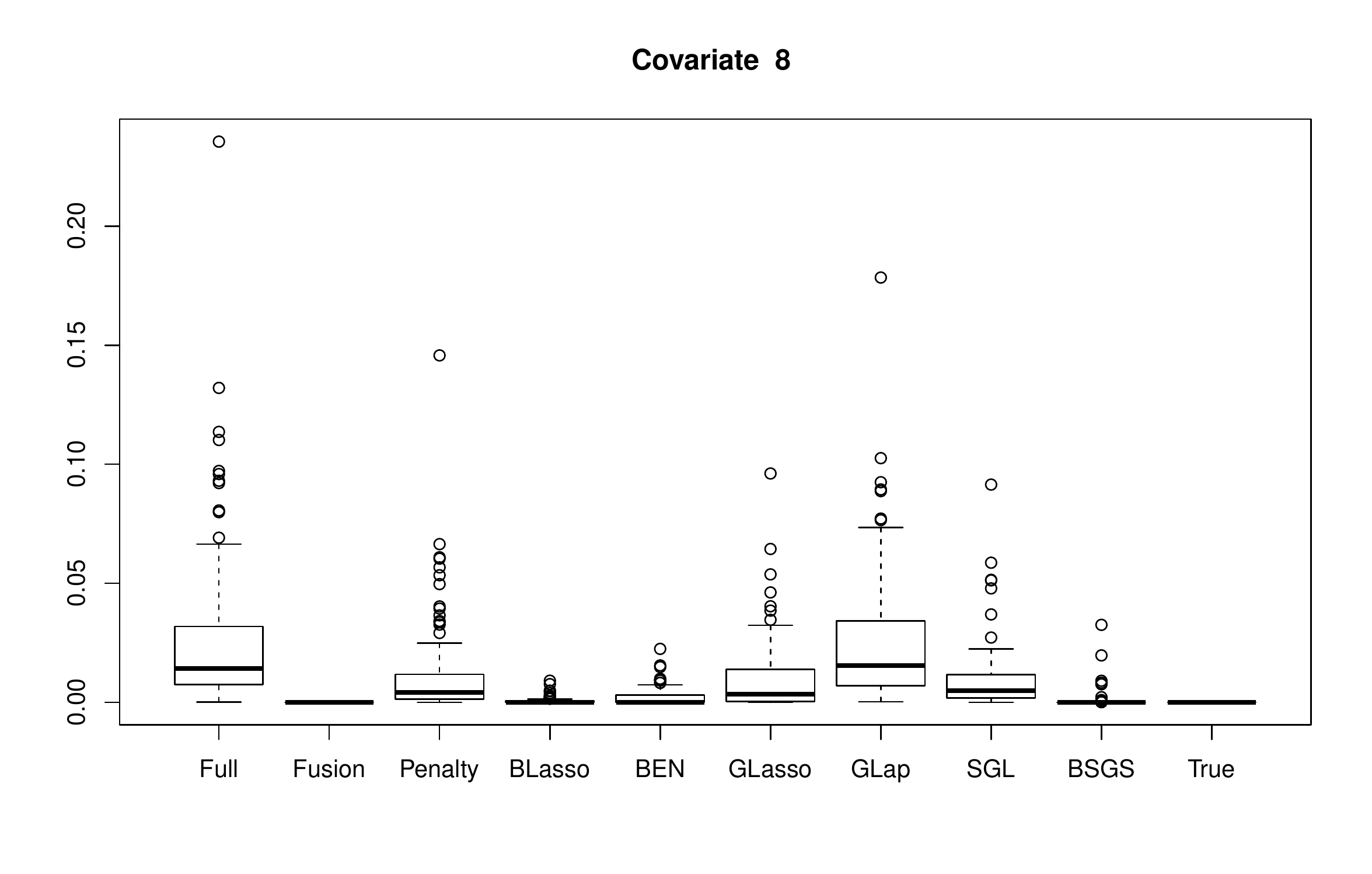}	
\caption{\label{fig:sim_mse} Simulation study: MSE for  ordinal (Covariates 1 - 4)  and nominal covariates (Covariates  5 - 8) in 100 simulated data sets. Covariates  2, 4, 6, and 8 (right panel) have no effect on the response. }						
\end{figure}

We first compare  the different methods with respect to estimation of the regression effects. 
Figure \ref{fig:sim_mse} shows  boxplots of  the mean squared estimation error (MSE), which is defined
 for data set $i$  and covariate $h$ as
 $$MSE_h^{(i)} = \frac{1}{c_h}\sum_{k=1}^{c_h}(\hat{\beta}_{h,k}^{(i)}-\beta_{h,k})^2.$$

No method outperforms the  others   consistently in all data sets for all covariates.  The mean MSE (averaged over all 100 data sets) is lower for  Bayesian effect fusion  (\emph{Effect Fusion}) than in  the model \emph{Full}   and only slightly higher than in the model  \emph{True}  for all covariates.  \emph{Effect Fusion} outperforms all other methods  with respect to the mean MSE for the ordinal covariates 1 and 3 and for  all  nominal covariates (5 - 8). However, due to high MSEs in some data sets it  is outperformed    by \emph{BLasso} and \emph{BSGS}   for  the ordinal covariates  2 and 4 (both with no effect on the response) and  for covariate 4 also  by  \emph{BEN}. 
Though  overall performance of \emph{Effect Fusion} is  very  good, for single data sets  the MSE can be even higher than for the model \emph{Full}. This occurs  when levels with actually different effects are fused, e.g. for some of the 8 levels of  the nominal covariate 5. 

The frequentist method for effect fusion,
\emph{Penalty}, which uses a global penalty parameter across all covariates  yields a lower MSE than the model \emph{Full}  for covariates with no effect and for ordinal covariates, but  only small improvements  for nominal covariates. 

 \emph{BLasso}  which  aims at shrinkage of  effects  to zero performs very well for covariates with no effect but also good for the other covariates 1, 3, 5 and 7. The performance of \emph{BEN} is similar, however slightly worse than  that  of \emph{BLasso} for all covariates.  As expected,  \emph{GLasso} which aims at sparsity at the group level  performs  well for covariates with no effect but   MSEs are similar to  those of the model \emph{Full} for covariates with non-zero effects. \emph{GLap}  which is  designed for a different goal yields small improvements  for covariates with no effect  compared to the model \emph{Full} and performs similar for covariates with non-zero effects. Finally,  both  \emph{SGL} and \emph{BSGS}, which aim  at sparsity at the group level as well as within groups of effects  outperform  \emph{Full}  for  covariates with no effect and  to a smaller extent also for the nominal covariates 5 and 7,  where some  level effects are zero. However,  \emph{SGL}   performs worse then \emph{Full} for  covariates 1 and 3,  where  the ordinal structure   is not taken into account.   \emph{BSGS} performs almost as good as  \emph{True} and similar to \emph{Effect Fusion} (slightly better for covariates 2 and 4, slightly worse for covariates 6 and 8) for covariates with no effects. For nominal  and ordinal covariates with non-zero effects it is outperformed by \emph{Effect Fusion} but similar as  the two other Bayesian methods  \emph{BLasso} and  \emph{BEN}. 
 

  Also model averaged estimates (not shown in  Figure \ref{fig:sim_mse}), which are  obtained as
  posterior mean estimates  from the first MCMC run
  under  the effect fusion prior,  perform very well with respect to MSE.

To evaluate the predictive performance of Bayesian effect fusion, we generate a new sample of $ n^*=500$ observations $ z_j,\; j=1,\dots, n^*$  from the linear regression model (\ref{model1}) with fixed regressors  $\tilde{\xv}_j$ and  the same parameters as in the simulated data sets. Predictions for these new observations are computed using the estimates from each of the original data sets as  $\hat{z}_j^{(i)}= \tilde\xv_j \hat{\betav}^{(i)}, i=1,\dots, 100$. 

\begin{figure}[h!]
\centering
	\includegraphics[scale=0.35]{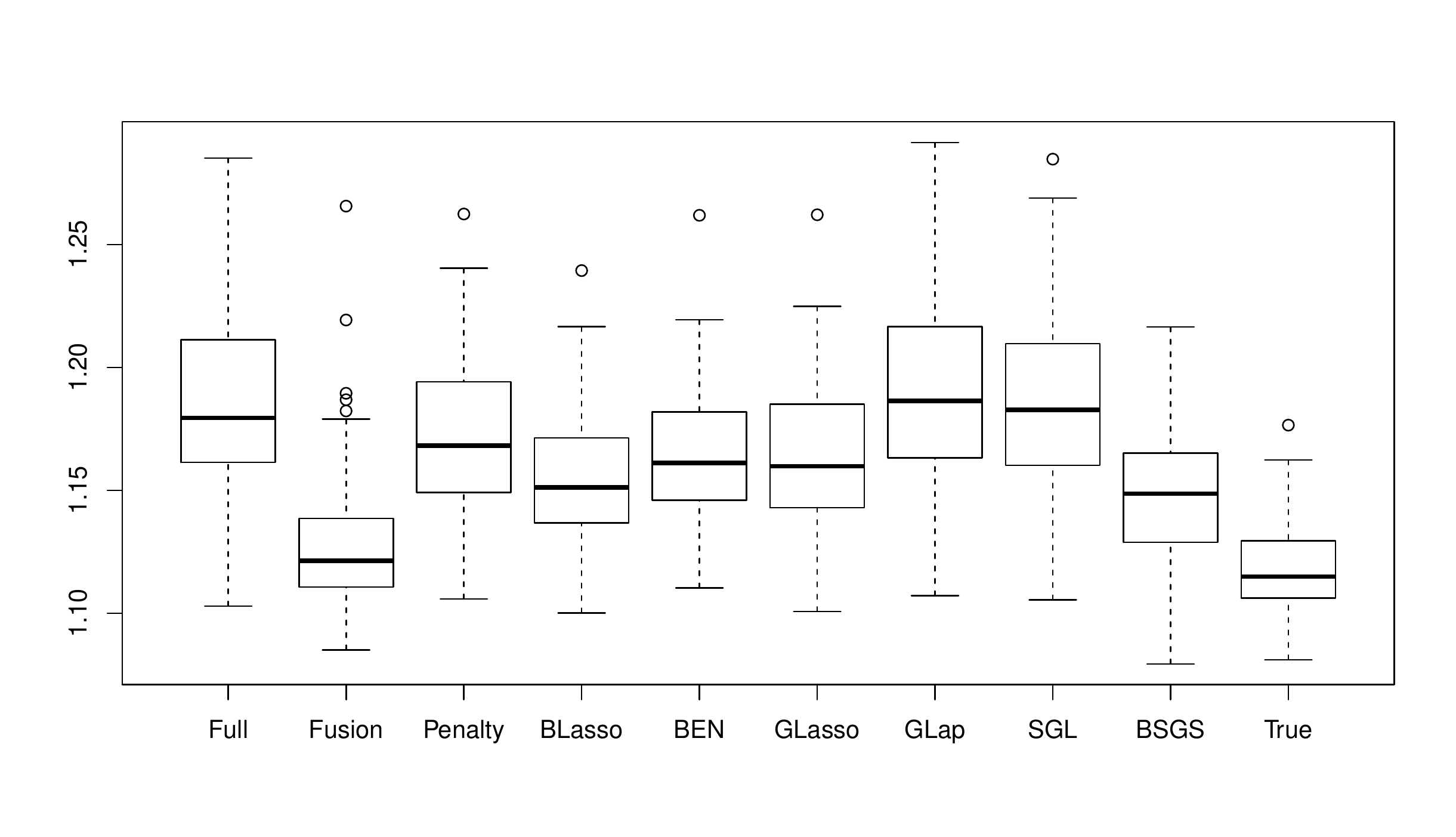}
	\caption{Simulation study: MSPE of 500 new observations. Predictions are based on the estimates from 100  simulated data sets.}
\label{res:mse_pred}
\end{figure}

The mean squared prediction errors (MSPE)  defined  for each data set as 
$$\text{MSPE}^{(i)}= \frac{1}{n^*} \sum_{j=1}^{n^*}( z_j - \hat{z}_j^{(i)})^2, \quad i=1,\dots,  100 $$
are shown in Figure \ref{res:mse_pred}. The  predictive performance of Bayesian effect fusion  is  almost as good as for the model \emph{True} with correctly fused effects,   and is considerably better for than all competing methods in most data sets. The  Bayesian methods \emph{BLasso} and  \emph{BSGS}   perform similar with respect to the mean MPSE (averaged over 100 data sets) and slightly better than \emph{Penalty}, \emph{BEN} and \emph{GLasso}. Finally, MSPEs for \emph{SGL} and \emph{GLap}  are  similar to  those of model \emph{Full}. 

Finally, to evaluate and compare the  performance of the methods  with respect  to model selection, we report for each covariate the true positive rate (TPR), the true negative rate (TNR), the positive predictive value (PPV) and the negative predictive value (NPV), see  Appendix \ref{app:sim_measures} for detailed definitions. If fusion is completely correct,  all four values are equal to 100\% but  TPR and PPV are not defined for covariates where all effects are zero.
For the effect fusion prior we perform model selection as described in Section \ref{sec:model_sel},  for the  other methods we  consider 
two level effects as identical if the  posterior mean of their difference  is smaller or equal to 0.01. Tables \ref{tab:model_res_ord}  and  \ref{tab:model_res_nom} report  the averages of these statistics in  the 100 simulated data sets for each covariate seperately.
 Bayesian effect fusion which  clearly outperforms all other methods with respect to  identifying categories with the same effect  
 with averaged TNR  higher than 95\% for all covariates. This comes at the cost of occasionally missing a non-zero effect difference and hence  an average TPR slightly lower than 100\%.

\begin{table}[!ht]
	\centering
	\footnotesize
	\begin{tabular}{clcccc}
		\hline
		Covariate & Method & TPR & TNR & PPV & NPV \\ 
		\hline
		1 & Fusion  & 99.7 & 95.3 & 95.3 & 99.8 \\ 
		&  Penalty & 100 & 18.8 & 48.8 & 100 \\ 
		&  BLasso  & 100 & 6.8 & 44.8 & 100 \\ 
		&  BEN     & 100 & 19.2 & 48.5 & 100 \\ 
		&  GLasso  & 100 & 2.2 & 43.5 & 100 \\ 
		& GLap     & 100 & 2.5 & 43.6 & 100 \\
		& SGL & 100 & 8.7 & 45.4 & 100 \\ 
		& BSGS    & 100  & 12.5 & 46.5 & 100 \\ 
		\hline
		2 & Fusion  &   -   & 98.3 &   -   & 100 \\  
		&  Penalty & - & 21.3 & - & 100 \\ 
		&  BLasso  & - & 24.1 & - & 100 \\ 
		&  BEN 	& - & 53.9 & - & 100 \\ 
		&  GLasso  & - & 21.7 & - & 100 \\ 
		& GLap     & - & 2.7 & - & 100 \\
		& SGL & - & 17.3 & - & 100 \\ 
		& BSGS    &  -    & 83.7 &    -   & 100 \\ 
		\hline
		3 & Fusion  & 100  & 99.0 & 99.3 & 100 \\ 
		&  Penalty & 100 & 24.5 & 44.8 & 100 \\ 
		&  BLasso  & 100 & 25.0 & 43.7 & 100 \\ 
		&  BEN     & 100 & 46.5 & 52.5 & 100 \\ 
		&  GLasso  & 100 & 10.5 & 36.8 & 100 \\ 
		& GLap     & 100 & 6.5  & 35.5 & 100 \\
		& SGL & 100 & 17.5 & 39.5 & 100 \\ 
		& BSGS    & 100  & 27.5 & 43.8 & 100 \\ 
		\hline
		4 & Fusion  &  -    & 99.0 &   -    & 100 \\ 
		&  Penalty & - & 21.3 & - & 100 \\ 
		&  BLasso  & - & 48.3 & - & 100 \\ 
		&  BEN 	& - & 61.3 & - & 100 \\ 
		&  GLasso  & - & 34.0 & - & 100 \\ 
		& GLap     & - & 7.3 & - & 100 \\
		& SGL & - & 18.0 &  & 100 \\
		& BSGS    &  -    & 81.7 &    -   & 100 \\ 
		\hline
	\end{tabular}
	\caption{Simulation study: Model selection results for ordinal covariates. TPR, TNR, PPV and NPV are averaged over 100  data sets.} 
	\label{tab:model_res_ord}
\end{table}

\begin{table}[!ht]
	\centering
	\footnotesize
	\begin{tabular}{clcccc}
		\hline
		Covariate & Method & TPR & TNR & PPV & NPV \\ 
		\hline  
		5 & Fusion  & 99.1 & 98.8 & 99.5 & 98.5 \\ 
		&  Penalty & 100 & 17.2  & 75.3 & 100 \\ 
		&  BLasso  & 100 & 6.8   & 72.9 & 100 \\ 
		&  BEN     & 99.9 & 12.0 & 74.0 & 99.5 \\ 
		&  GLasso  & 100 & 4.2   & 72.3 & 100 \\ 
		& GLap     & 100 & 4.0   & 72.3 & 100 \\
		& SGL & 100 & 7.1 & 73.0 & 100 \\
		& BSGS    & 100  & 7.6  & 73.1 & 100 \\ 
		\hline
		6 & Fusion  &  -    & 100  &   -    & 100 \\ 
		&  Penalty & - & 51.4 & - & 100 \\ 
		&  BLasso  & - & 27.5 & - & 100 \\ 
		&  BEN 	& - & 54.5 & - & 100 \\ 
		&  GLasso  & - & 21.3 & - & 100 \\ 
		& GLap     & - & 3.9 & - & 100 \\
		& SGL  &   - & 19.5 &  & 100 \\ 
		& BSGS    &    -  & 84.2 &   -    & 100 \\ 
		\hline
		7 & Fusion  & 100  & 99.5 & 99.8 & 100 \\  
		&  Penalty & 100 & 17.5 & 71.6 & 100 \\ 
		&  BLasso  & 100 & 22.5 & 72.7 & 100 \\ 
		&  BEN     & 100 & 43.0 & 78.3 & 100 \\ 
		&  GLasso  & 100 & 5.5  & 68.1 & 100 \\
		& GLap     & 100 & 4.5  & 67.9 & 100 \\
		& SGL & 100 & 15.0 & 70.7 & 100 \\ 
		& BSGS    & 100  & 32.0 & 75.6 & 100 \\ 
		\hline 
		8 & Fusion  &   -   & 100  &    -   & 100 \\ 
		&  Penalty & - & 27.5 & - & 100 \\ 
		&  BLasso  & - & 47.8 & - & 100 \\ 
		&  BEN     & - & 70.2 & - & 100 \\ 
		&  GLasso  & - & 35.3 & - & 100 \\  
		& GLap     & - & 4.7 & - & 100 \\
		& SGL  & - & 19.3 &  & 100 \\
		& BSGS    &   -   & 85.3 &   -    & 100 \\ 
		\hline
	\end{tabular}
	\caption{Simulation study: Model selection results for nominal covariates. TPR, TNR, PPV and NPV are averaged over 100  data sets.} 
	\label{tab:model_res_nom}
\end{table}

\subsection{Influence of hyperparameters} \label{sec:sim_hyp}

In this section, we investigate  the sensitivity of model selection  under the effect fusion prior with respect  to the  hyperparameters.   While false positives result in a loss of estimation efficiency,  false negatives  will yield biased effect estimates and poor predictive performance,  and hence  the goal is to avoid \emph{false negatives} while keeping \emph{false positives}
at a moderate level.  We  report  false negative rates, $\text{FNR} = 1 - \text{TPR}$ and  false positive rates  $\text{FPR} = 1 - \text{TNR}$ for  various values of  $G_0$  and fixed  $r=20000$ in Table \ref{tab:sim_G} and for various values of  $r$  and  fixed $G_0=20$ in  Table \ref{tab:sim_r}.
Results  in Table \ref{tab:sim_G}  indicate that increasing $G_0$  from $0.2$ to $200$  has little effect on FNR but yields  lower FPR  for ordinal predictors (covariates 1  --  4),   and  has little effect on FPR but leads to higher FNR for nominal predictors (covariates 5  --  8).  We conclude  that  for nominal predictors $G_0=2$ is a good choice which allows to detect also small effect differences  whereas for  ordinal predictors we suggest to choose a larger value for $G_0$, e.g.~$G_0=20$. An exponential hyperprior  with $\E(G_0)=2$ performs  similar to a fixed value of  $G_0=2$ with  slightly lower FNR and  FPR for nominal covariates but  higher FPR for ordinal covariates.

 Table  \ref{tab:sim_r}   reports  FNR and FPR   for  values of the precision ratio $r$ from $2 \cdot 10^2$ to $2 \cdot 10^5$.  Whereas for ordinal predictors 
 FNR and FPR change only little with $r$, for nominal covariates  low values of  $r$  encourage too much fusion of effects and hence result in a high FNR.  Both FNR and FPR are small  for $r = 2\cdot 10^4$ and $r = 2 \cdot10^5$. As  MCMC  mixing is adequate for both values, 
we  suggest  to choose $r$ in this range.

\begin{table}[!ht]
\centering
\small
\begin{tabular}{|c|cc|cc|cc|cc|cc|}
  \hline
 & \multicolumn{2}{c|}{$G_0=0.2$} &\multicolumn{2}{c|}{$G_0=2$} &\multicolumn{2}{c|}{$G_0=20$} &\multicolumn{2}{c|}{$G_0=200$} & \multicolumn{2}{c|}{$G_0 \sim \Exp{2}$} \\
h  & $FNR_h$ & $FPR_h$ & $FNR_h$ & $FPR_h$ & $FNR_h$ & $FPR_h$  & $FNR_h$ & $FPR_h$ & $FNR_h$ & $FPR_h$ \\ 
  \hline
  1 &  0.0 & 13.0 & 0.3 & 10.5 & 0.3 & 4.7 & 0.7 & 1.7 & 0.3 & 9.0 \\
  2 &   -     & 18.9 &  -     & 7.4 & -       & 1.7 & -        & 0.3 &  -         & 11.7 \\
  3 &  0.0 & 11.0 & 0.0 & 7.5 & 0.0 & 1.0 & 0.0 & 0.5  & 0.0 & 5.0 \\
  4 &   -    & 14.3 & -        & 5.0 &   -      & 1.0 &   -      & 0.3 &   -     & 8.7 \\
  5 &  0.4 & 1.2   & 0.7 & 1.0 & 0.9   & 1.2 & 81.0 & 0.0 & 0.5 & 0.9 \\
  6 &   -     & 0.0  &  -       & 0.0 &  -       & 0.0 & -       & 0.0  &   -     & 0.0 \\
  7 &  0.0 & 3.0  & 0.0 & 2.0 & 0.0   & 0.5 & 0.0 & 0.0  & 0.0 &1.5  \\
  8 &   -     & 0.0 & -        & 0.0 &  -       & 0.0 & -        & 0.0  &  -     & 0.0 \\
   \hline
\end{tabular}
\caption{Simulation study: Model selection results for $r=20,000$ and various values of $G_0$} 
\label{tab:sim_G}
\end{table}

\begin{table}[!ht]
\centering
\small
\begin{tabular}{|c|cc|cc|cc|cc|}
  \hline
 & \multicolumn{2}{c|}{$r=200$} &\multicolumn{2}{c|}{$r=2,000$} &\multicolumn{2}{c|}{$r=20,000$} &\multicolumn{2}{c|}{$r=200,000$}  \\
h  & $FNR_h$ & $FPR_h$ & $FNR_h$ & $FPR_h$ & $FNR_h$ & $FPR_h$  & $FNR_h$ & $FPR_h$   \\ 
  \hline
  1 & 0.3 & 2.7 & 0.3 & 4.2 & 0.3 & 4.7 & 0.3 & 3.0 \\
  2 &   -      & 0.9 & -       & 1.4 &  -       & 1.7 &  -       & 1.7 \\ 
  3 & 0.0 & 0.5 & 0.0 & 1.5 & 0.0 & 1.0 & 0.0 & 2.0 \\
  4 &  -       & 0.3 & -        & 0.3 &  -      & 1.0 & -        & 1.7 \\
  5 & 100.0 & 0.0 & 92.0 & 0.0 & 0.9 & 1.2 & 0.8 & 1.9 \\
  6 &   -      & 0.0 & -        & 0.0 &   -      & 0.0 & -        & 0.6  \\
  7 & 0.0 & 0.0 & 0.0 & 0.0 & 0.0 & 0.5 & 0.0 & 0.5 \\
  8 &   -     & 0.0 & -        & 0.0 &  -       & 0.0 & -        & 0.0 \\
   \hline
\end{tabular}
\caption{Simulation study: Model selection results for $G_0=20$ and various values of  $r$} 
\label{tab:sim_r}
\end{table}

\section{Real data example} \label{sec:application}

As an  illustration of  Bayesian effect fusion  on real data,  we model  contributions to private retirement pension in Austria. The data were collected in the European household survey EU-SILC (SILC = Survey on Income and Living Conditions) 2010 in Austria. We use a linear regression model to analyse the effects of socio-demographic variables on the (log-transformed) annual contributions to private retirement pensions. As potential regressors we consider  {\tt gender} (binary, 1=female/0=male), {\tt age group} (ordinal with eleven levels), {\tt child} in household (binary, 1=yes/0=no), {\tt income class} (in quartiles of the total data set, i.e.~ordinal with four  levels), {\tt  federal state} of residence in Austria (nominal with  nine levels), highest attained level of {\tt education} (nominal with ten levels)  and {\tt employment status} (nominal with four levels). We restrict the analysis to observations without missing values in either regressors  or response   and a  minimum annual contribution of EUR 100. Hence, the final data set used for our analysis comprises data of 3077 persons.

We standardised the response and fit a regression model including all potential covariates. Results  reported in Table \ref{tab:silc_full} indicate that  several levels of covariate  {\tt education} have a similar effect and  most level effects of {\tt federal state} are close to zero, which suggests that a sparser model might be adequate for these data.
\begin{table}[!ht]
	\centering
	\scriptsize
	\begin{tabular}{lcc|ccc}
		\hline
		& Full model & 95\% && Selected model & 95\%  \\
		& Posterior mean  & HPD interval && Posterior mean & HPD interval \\ 
		\hline  
		Intercept & -1.15 & (-1.38 -- -0.91) && -1.10 & (-1.31 -- -0.90) \\ 
		\hline
		{\bf Age} & & & &&  \\
		20-25 & 0.20 & (-0.03 -- 0.42) & \rdelim\}{2}{0.5mm} & 0.33 & (0.14 -- 0.52)  \\ 
		25-30 & 0.36 & (0.15 -- 0.57)  & &   &  \\ 
		30-35 & 0.60 & (0.40 -- 0.80)  & & 0.65  & (0.45 -- 0.86) \\ 
		35-40 & 0.74 & (0.53 -- 0.95)  & \rdelim\}{2}{0.5mm} &   &  \\ 
		40-45 & 0.80 & (0.60 -- 1.00)   & & 0.82  & (0.62 -- 1.01) \\ 
		45-50 & 0.90 & (0.70 -- 1.10)  & & 0.93  & (0.74 --  1.13) \\ 
		50-55 & 1.01 & (0.80 -- 1.23)  & \rdelim\}{4}{0.5mm} & & \\ 
		55-60 & 1.06 & (0.81 -- 1.30)  & &  & \\ 
		60-65 & 1.23  & (0.83 -- 1.62)  & &  1.05  &  (0.85 -- 1.24)  \\ 
		$>$ 65 & 0.67 & (0.14 -- 1.22)  & & & \\ 
		\hline
		{\bf Female} & -0.24 & (-0.32 -- -0.17)  & &  -0.25  & (-0.31 -- -0.18) \\ 
		\hline
		{\bf Child} &  0.00 &  (-0.07 -- 0.07)  &&  -  & -  \\ 
		\hline
		{\bf Income} & & & & & \\
		2nd quartile & 0.20 & (0.07 -- 0.32)  & \rdelim\}{2}{0.5mm} &     &  \\ 
		3rd quartile & 0.25 & (0.13 -- 0.37)  &&   0.23  &  (0.12 -- 0.35) \\ 
		4th
		 quartile & 0.52 & (0.40 -- 0.64) & &  0.54  & (0.42 -- 0.66) \\ 
		\hline
		{\bf Federal State} & & & & & \\
		Carinthia 	& -0.16 &  (-0.30 -- -0.01) &  & -  & -\\ 
		Lower Austria & 0.06 &  (-0.03 -- 0.16)  & & -  & -\\ 
		Burgenland 	& -0.03  & (-0.21 -- 0.15)&  &  -  & -\\ 
		Salzburg 		& 0.15 &   (0.01 -- 0.30) &  & -  & -\\ 
		Styria 		& 0.01 & (-0.10 -- 0.13) &  &  -  & -\\ 
		Tyrol 		& 0.08 & (-0.06 -- 0.20) &  & -  &- \\ 
		Vorarlberg 	& 0.02 & (-0.15 -- 0.19) & &  -  & -\\ 
		Vienna 		& 0.00 & (-0.11 -- 0.10) &  & -  & - \\ 
		\hline
		{\bf Education} & & &&  & \\
		Apprenticeship, trainee          & 0.09 & (-0.04 -- 0.23) &  & 0.00  & - \\ 
		Master craftman's diploma        & 0.24 & (0.06 -- 0.43) & \rdelim\}{10}{0.5mm} &    &   \\ 
		Nurse's training school          & 0.22 &  (-0.04 -- 0.47) &  &   &  \\ 
		Other vocational school  &&&&& \\
		(medium level)                   & 0.26 &  (0.11 -- 0.42) &  &   &  \\
		Academic secondary school &&&&& \\
		(upper level)                    & 0.22 &  (0.05 -- 0.37) &  &  0.21  & (0.14 -- 0.28) \\ 
		College for higher vocational &&&&& \\
		education                           & 0.28 &  (0.12 -- 0.43) &  &   & \\ 
		Vocational school for apprentices   & 0.29 &  (0.06 -- 0.51) &  &   &  \\ 
		University, academy: first degree   & 0.35 &   (0.20 -- 0.49)&  &   &  \\ 
		University: doctoral studies        & 1.12 &  (0.85 -- 1.37) &  & 1.03  & (0.80 --  1.26) \\ 
		\hline
		{\bf Employment status} &&& && \\
		Unemployed 	& -0.10 & (-0.34 -- 0.12) & & -  & - \\ 
		Retired 		& -0.15 & (-0.37 -- 0.07)  & &  -  & - \\ 
		Not-working &&& && \\
		(other reason) & 0.01 & (-0.11 -- 0.14)  & & -  & - \\ 
		\hline
	\end{tabular}
	\caption{\label{tab:silc_full} EU-SILC data: Posterior means and 95\% HPD intervals of regression effects in the full  and in the selected model.}
\end{table}

To specify the  effect fusion prior, we chose the  hyperparameters with  $r=50000$,  $g_{h0} = 5$, $G_{h0} = 2$  for nominal and $G_{h0} = 20$ for ordinal predictors and used the improper prior $p(\sigma^2)=\frac{1}{\sigma^2}$. MCMC  was run for 50000 iterations after a burn-in of 30000 with the first 500 draws of the burnin drawn from the unrestricted model where all elements of  $\deltav$ are 1. The median 
of the integrated autocorrelation times over all regression effects  was  11.9 (range 1 - 104) and  integrated  autocorrelation times  were lower than 30 for effects of all covariates except age. We ran several chains from different starting values which yielded essentially the same results.  

\begin{figure}[h!]
	\centering
	\includegraphics[scale=0.4]{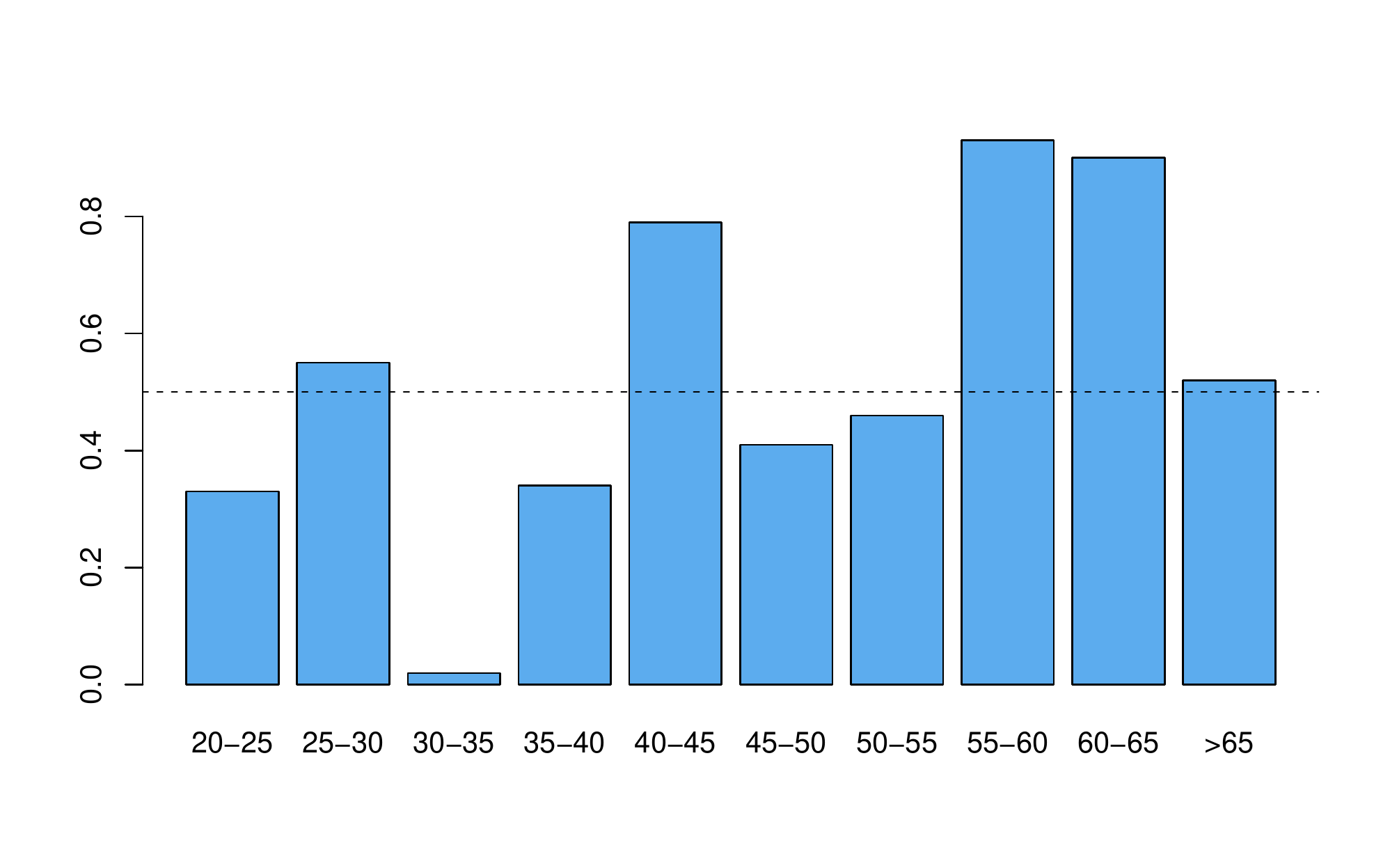}
	\caption{\label{fig:fusion_age} EU-SILC data, covariate {\tt age group}: estimated  probabilities for fusion with preceding level}
\end{figure}

Figure  \ref{fig:fusion_age}  shows the estimated posterior means of the pairwise fusion probabilities $1-\hat{\pi}_{h, k k-1}$  for the ordinal covariate {\tt age group}. The estimated
  fusion probability  is higher than 0.5 (dotted line) for five levels,  which indicates  that  age categories could be fused as follows:
  categories  $20-25$ and $25-30$  to $20-30$, categories   $35-40$ and $40-45$ to $35-45$ and finally  
  the three categories $50-55$, $55-60$ and $60-65$  to a new category  $50-65$.

\begin{figure}[h!]
	\centering
	\includegraphics[scale=0.5]{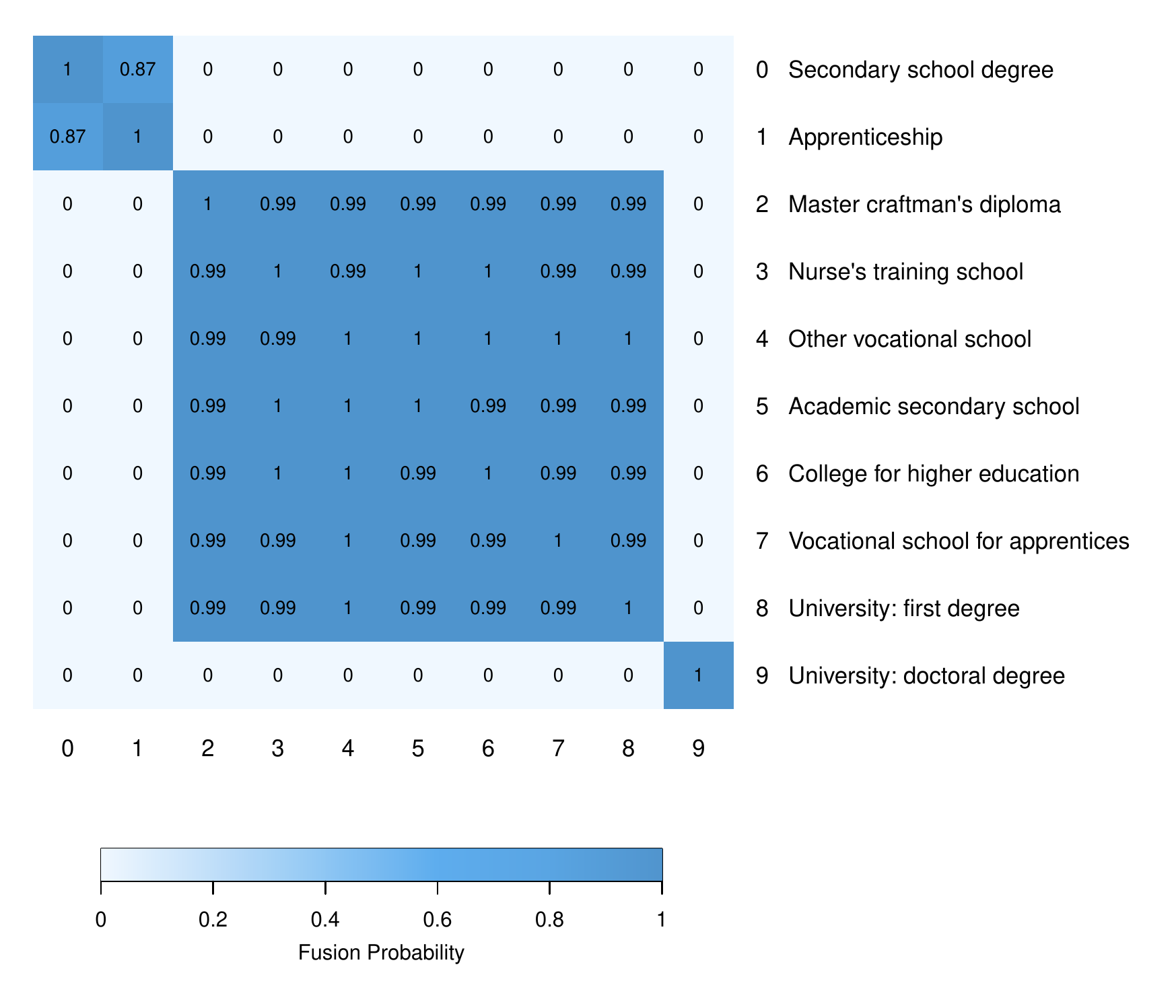}
\caption{\label{fig:fusion_education} EU-SILC data, covariate {\tt education}: Estimated pairwise fusion probabilities of level effects (baseline: category 0). Darker colours indicate higher fusion probabilities.}
\end{figure}

For a  nominal covariate  we  suggest to visualize the pairwise fusion probabilities  in a heatmap.  Figure \ref{fig:fusion_education} shows the  corresponding heatmap for the covariate {\tt education}, where all pairwise fusion probabilities are displayed. Darker colours indicate  higher fusion probabilities and values  in the diagonal (which represent fusion probability of a category with itself) are always one.
 Obviously, only three  levels  are required to capture the effect of education (secondary school and  apprenticeship; doctoral degree; all remaining levels) and thus   the number of effects to be estimated reduces from  nine to two.


Based on the estimated pairwise fusion probabilities we  select the final model  for all covariates as described in Section \ref{sec:model_sel}. Covariates {\tt child, federal state} and {\tt employment status} are  completely excluded from the model, whereas some of the  levels are fused for covariates {\tt age group, income class} and {\tt education}. Thus, the final model has only 11 regression effects  compared to 35  in the full model. Results from a refit of the selected model using flat priors are reported in the right panel of Table  \ref{tab:silc_full}. The posterior mean of the error variance, $\hat{\sigma}^2=0.829$, is almost identical to that of the full model, where $\hat{\sigma}^2=0.826$.
         
   
\section{Conclusion} \label{sec:conclusion}     
            
In this paper, we present a method that allows  for  sparse modelling of  the effects of categorical covariates in  Bayesian regression models. Sparsity is achieved by excluding irrelevant predictors and  by fusing levels which have essentially the same effect on the response. To encourage effect fusion, we propose a finite mixture of  Normal prior distributions with component specific precision matrices,
 that allow for almost perfect or  almost zero  partial dependence of  level effects. Alternatively, this  prior can be derived by specifying  spike and slab prior distributions on all level effect contrasts associated with one covariate  and   taking   the  linear restrictions among them  into account.  The structure of   prior easily  allows to incorporate  prior information  that restricts direct fusion to specific  pairs of level effects. This property is of particular interest for ordinal covariates where fusion usually will be restricted to effects of subsequent levels.

Posterior inference  for all  model parameters is straightforward using MCMC methods. To select the final model we 
suggest to determine  the clustering of level effects which minimizes  Binder's loss
 based on the estimated posterior means of the pairwise fusion probabilities. Simulation results show  that the proposed method automatically  excludes irrelevant predictors  and outperforms competing methods  in terms of correct model selection, coefficient estimation as well as predictive performance.

The proposed method for Bayesian effect fusion is not restricted to  linear regression models with categorical predictors
 but can be applied in more general regression type models e.g.~generalised linear models, with a structured additive predictor that 
 also contains other types of effects, e.g. nonlinear  or spatial effects.  Implementation of effect fusion requires little adaption of an MCMC scheme for posterior sampling of any Bayesian regression type model with a Gaussian  prior on the regression effects as  only two  Gibbs sampling steps and the
update of the structure matrix have to be added in each MCMC sweep.

 A certain drawback of the method is that all pairwise effect differences have to be determined and classified in each MCMC sampling step  to construct  the prior covariance matrix and hence  the computational effort can be prohibitive for nominal covariates with a large number of levels. If we consider each combination of spike and slab distribution on effect differences as a different model,  then for a nominal covariate with $c+1$ levels  the  procedure  searches a space of  $2^{c(c+1)/2}$ different models. This number is much larger than the number of possible clusterings, which is  given by the Bell number  $B(c+1)$ of order  $c$.  A  search in the model space  of possible clusterings could be performed by employing  a reversible jump MCMC algorithm  as in \cite{del-tar:mod}.
Another option to determine a model with eventually fused level effects was  recently proposed in \cite{mal-etal:eff}. The authors use model-based clustering methods and specify a finite mixture prior with many spiky components on the level effects. The    prior on the mixture weights encourages empty components and thus some level effects will be  assigned to the same mixture component. Due to the spikyness of the components, almost identical effects are  assigned to the same component which suggests to fuse effects according to the clustering solution. This approach works well for nominal covariates but does not easily allow incorporation of fusion restrictions. 

\vspace{1 cm}
{\bf Acknowledgement:} This work was financially supported by the Austrian Science Fund (FWF) via the research project number P25850 'Sparse Bayesian modelling for categorical predictors'.

\appendix
 \section{Properties of the  effect fusion prior}   \label{app:fusion}
 
  
  \subsection{Properties of the structure matrix }   \label{app:pri_pd}
  For simplicity of notation, we drop dependence of the structure matrix  $\Qv(\deltav)$ on $\deltav$  in the following.
 $\Qv$ is symmetric by definition. To show  its positive definiteness  we consider the quadratic form $\xv' \Qv \xv$
   where $\xv$ is a non-zero vector of dimension $c$. 
     
  Obviously, we have 
   \begin{align*} \xv' \Qv \xv & = \sum_{k =1}^c \sum_{j =1}^c x_k x_j q_{kj}= \\
   &=  \sum_{k =1}^c x_k^2 \Big(\sum _{j\neq k} \kappa_{kj} \Big)- \sum_{k =1}^c \sum\limits_{\substack{j =1 \\ j \neq k}} ^c  x_k x_j \kappa_{kj}
   = \\
   &=   \sum_{k =1}^c   x_k^2  \kappa_{k0} +  \sum_{k =1}^c \sum\limits_{\substack{j =1 \\ j \neq k}} ^c x_k^2  \kappa_{kj}- 
   \sum_{k =1}^c \sum\limits_{\substack{j =1 \\ j \neq k}} ^c  x_k x_j \kappa_{kj} =\\
   &=   \sum_{k =1}^c   x_k^2  \kappa_{k0} + \sum_{k =2}^c  \sum_{j=1}^{k-1} ( x_k - x_j)^2  \kappa_{kj} >0
  \end{align*}
  if $\xv \neq \zerov$, where  $\zerov$ is a vector of zeros.

  
  \subsection{Spike and slab priors on effect differences } \label{app:pri_diff}
  We show that the prior on the regression effects $\betav$ given  in equation (\ref{pri1})  with  structure matrix $\Qv(\deltav)$ specified in  equation (\ref{eq:prec}) corresponds to  a prior, where  independent spike and slab priors  are specified on all effect differences and then the  linear restrictions implied by their definition are corrected for.
  
  Spike and slab priors on the effect differences $\theta_{kj}=\beta_k -\beta_j$ can be  specified for  all pairs $k,j$ with $0 \le j < k \le c$ as  
  \begin{align*}
  	\theta_{kj}|\delta_{kj},\tau^2 & \sim \delta_{kj}\Normal{0,  \tau^2 \gamma} + (1-\delta_{kj})  \Normal{0,  \frac{1}{r} \tau^2  \gamma} \\
     	\tau^2 & \sim \mathcal{G}^{-1}(g_{0}, G_{0}). 
  \end{align*}

  Conditional on the hyperparameters, we write  the  prior  on  the  effect difference $\theta_{kj}$    more compactly as
  $$\theta_{kj} \sim \Normal{0, \frac{\tau^2 }{\kappa_{kj}} \gamma}.$$
  
  We subsume all effect differences $\theta_{kj}$ in the $d \times  1 $ vector $\thetav$ so that the subvector $\thetav_0$ of the first $c$ elements equals
   $\betav$, i.e. $\thetav_0=(\theta_{10} , \dots, \theta_{c0})'=\betav$,  and  partition $\thetav$ accordingly in $\thetav=(\thetav_0', \thetav_{\setminus 0}')'$.
  We write the linear restrictions
  $$\theta_{kj}=\beta_k-\beta_j =\theta_{k0} -\theta_{j0} $$
    in matrix form as 
   $ \Uv \thetav_0 - \thetav_{\setminus 0} = \zerov$, where $\Uv$ is the appropriate coefficient matrix. 
  
  The distribution of a Normal vector $\thetav \sim \Normal{\zerov, \Sigmav}$ under 
   the  linear restriction   $\Rv \thetav=\zerov$
  is again Normal with moments 
   \begin{align*}
  \E(\thetav| \Rv \thetav=\zerov ) & =\zerov \label{cond_mean}\\
  \Cov(\thetav| \Rv \thetav=\zerov) &=\Sigmav - \Sigmav  \Rv' (\Rv \Sigmav \Rv')^{-1} \Rv \Sigmav, 
  \end{align*}
  see \cite{rue-hel:gau}, p. 37.
  
 To determine  the covariance matrix $\Omegav= \Cov(\thetav| \Rv \thetav=\zerov)$  with
    $\Rv=\begin{pmatrix}  \Uv & -\Iv \end{pmatrix}$, where $\Iv$ is  the identity  matrix of dimension $d-c$,  we partition also $\Sigmav$ as   
  $$\Sigmav=\gamma \tau^2  \Deltav=   \gamma\tau^2 \begin{pmatrix}
  			\Deltav_{0} & \\
  			& \Deltav_{\setminus 0}= \end{pmatrix}= \gamma \tau^2 \begin{pmatrix} \kappav_{0}^{-1} & \\ 	& \kappav_{\setminus 0}^{-1} \end{pmatrix}$$
  	where $\kappav_0=\diag(\kappa_{10},\dots, \kappa_{c,0})$ and 	 $\kappav_{\setminus 0}$ is the diagonal matrix with elements
  	$\kappa_{kj}$ where $  j\neq 0$.
 Thus  \begin{align*}\Sigmav \Rv' & = \gamma \tau^2 \begin{pmatrix}  \Deltav_0  \Uv' \\ -\Deltav_{\setminus 0}\end{pmatrix}\\
 (\Rv \Sigmav \Rv')^{-1}& =\frac{1}{\gamma\tau^2}
 (  \Uv\Deltav_{0} \Uv'+\Deltav_{\setminus 0})^{-1}=
\frac{1}{\gamma\tau^2} \Wv
 \end{align*}
 and  $\Omegav$ results as
  \begin{align*}
   \Omegav& = \gamma \tau^2\Big(\Deltav- \begin{pmatrix}  \Deltav_0  \Uv' \\ -\Deltav_{\setminus 0}\end{pmatrix} \Wv 
   \begin{pmatrix}  \Uv \Deltav_0 & -\Deltav_{\setminus 0}\end{pmatrix}\Big) = \\ \smallskip   
  & =  \gamma \tau^2  \left[ \begin{pmatrix}
  \Deltav_0 & \\
  & \Deltav_{\setminus 0}
  \end{pmatrix} - \begin{pmatrix}
  \Deltav_0  \Uv' \Wv  \Uv \Deltav_0 & -\Deltav_0  \Uv' \Wv \Deltav_{\setminus 0} \\
  - \Deltav_{\setminus 0} \Wv  \Uv \Deltav_0  & \Deltav_{\setminus 0} \Wv \Deltav_{\setminus 0}
  \end{pmatrix}\right].
  \end{align*}

  As $\Bv_0(\deltav,\tau^2)$      is the  upper left  $c \times c$ matrix of  $ \Omegav$, i.e.
     $$\Bv_0(\deltav,\tau^2)=\gamma \tau^2  \Big(\Deltav_ {0}-\Deltav_ {0} \Uv' \Wv  \Uv\Deltav_ {0}\Big),$$
  the  structure matrix $\Qv(\deltav)$  is    given as
  \begin{equation} \Qv(\deltav)  = (\Deltav_ {0}-\Deltav_ {0} \Uv' \Wv  \Uv\Deltav_ {0})^{-1}=\Deltav_{0}^{-1}+  \Uv' \Deltav_{\setminus 0}^{-1}  \Uv=
  \kappav_{0}+  \Uv' \kappav_{\setminus 0} \Uv, \label{eq:Qmat}\end{equation}
 where the second identity results from the Woodbury formula.
  
  Denoting by   $ \uv_{k}$   the k-th column of $ \Uv$,
  the off-diagonal elements of $\Qv(\deltav)$ are given as
  $$q_{kj}= \uv_{k}' \kappav_{\setminus 0}   \uv_{j}= -  \kappa_{kj},$$ 
since for each pair of columns $ \uv_{k}$ and $ \uv_{j}$   both vectors have a non-zero element   only in the row  corresponding to the linear restriction for $\theta_{kj}$, with  value   $1$ in one  and $-1$ in the other vector.  
   
  Finally,  for the  diagonal elements of  $\Qv(\deltav)$   we get 
  $$	q_{kk}  =\kappa_{k0}+   \uv_{k} '\kappav_{\setminus 0} \uv_{k} = \kappa_{k0}+\sum\limits_{\substack{j =1 \\ j \neq k}} ^c  \kappa_{kj}= \sum_{j \neq k}\kappa_{kj}.
  $$
  
   \subsection{Invariance with respect to the baseline  category}  \label{app:inv}
  To show invariance of the effect fusion prior with respect to the baseline category,  without loss of generality we   change the  baseline category  of a nominal predictor from $0$ to $c$. For simplicity of notation, we write $\Qv$ instead of  $\Qv(\deltav)$.
  
  Let $\tilde \betav=(\tilde \beta_1,\dots,\tilde \beta_{c-1},\tilde \beta_c)$ denote the regression effects with respect to category $c$, i.e.  $\tilde \beta_k=\beta_k-\beta_c$ for  $k=1\dots, c-1$,  and    $\tilde \beta_c= \beta_0-\beta_c=-\beta_c$, and by
 $\Av$  the transformation matrix, so that
 $$\tilde \betav=\Av\betav.$$
 
 The precision matrix of $\tilde \betav$ is then given as
 $$\Cov^{-1}(\tilde{\betav})= \tilde \Qv/(\gamma \tau^2)= (\Av')^{-1} \Qv \Av^{-1} /(\gamma \tau^2),$$
 where $ \tilde \Qv$ denotes the corresponding structure matrix.
We partition $\Av$ and $\Qv$ as
 $$ \Av=\begin{pmatrix} \Iv_{c-1} & -\ev_{c-1}\\
 \zerov_{c-1}' & -1 \end{pmatrix}  \quad \text{and} \quad \Qv=\begin{pmatrix}\Qv_{c-1} & \qv_{c}\\
 \qv'_c & q_{c c}\end{pmatrix},$$
 where $  \Iv_{c-1}$ is  the unit matrix and $ \zerov_{c-1}$ and $\ev_{c-1}$    are  vectors of  zeros and ones of dimension $c-1$, respectively. $\Qv_{c-1}$ is the submatrix of the first $c-1$ rows and columns   of $\Qv$ and  $q_{c,c}$ is its the last element.  We denote the k-th column  of $\Qv_{c-1}$ by   $\qv_{k}=(q_{1k},\dots, q_{c-1,k})'$  and correspondingly define $\qv_{c}=(q_{1c},\dots, q_{c-1,c})'$.

  Obviously   $ \Av^{-1}= \Av$  and therefore  the structure matrix for $\tilde \betav$ results as
\begin{align*} \tilde \Qv &=\begin{pmatrix}  \Iv_{c-1} & \zerov_{c-1}\\ -\ev'_{c-1} &-1 \end{pmatrix} \begin{pmatrix}\Qv_{c-1} & \qv_{c}\\
 \qv_{c}' & q_{c,c}\end{pmatrix} \begin{pmatrix} \Iv_{c-1} & -\ev_{c-1}\\
 \zerov_{c-1}' & -1 \end{pmatrix} = \\ 
 &=\begin{pmatrix}\Qv_{c-1} &-\Qv_{c-1}\ev_{c-1} -  \qv_{c}\\
-\ev_{c-1}' \Qv_{c-1} -  \qv'_{c} & \ev_{c-1}' \Qv_{c-1}\ev_{c-1}+\ev_{c-1}'  \qv_{c}+ \qv'_{c}\ev_{c-1} + q_{c,c} \end{pmatrix} .
 \end{align*}
 To determine the elements of the last column of this matrix, we note that for $k=1,\dots, c-1$,
     $$ \qv_k' \ev_{c-1} + q_{kc} =\sum\limits_{j =1}^{c-1} q_{kj} + q_{kc}= q_{kk}+  \sum\limits_{\substack{j =0 \\ j \neq k}}^c q_{kj}=     
          \sum\limits_{\substack{j =0 \\ j \neq k}}^c \kappa_{kj} -  \sum\limits_{\substack{j =1 \\ j \neq k}}^c \kappa_{kj}=\kappa_{k0}.$$
   Therefore we get
$$ -\Qv_{c-1}\ev_{c-1} - \qv_{c}= - (\kappa_{10} \dots, \kappa_{c-1,0})', $$  and  the lower left  element of $\tilde \Qv$ is equal to
$$\tilde q_{cc}= \ev_{c-1}'  (\kappa_{10} \dots,\kappa_{c-1,0}) + \kappa_{c0} =\sum_{k \neq 0} \kappa_{k0}. $$
Thus the  prior structure matrix  of the level effects under  reference  category  is  $c$ is given as
$$\tilde \Qv= \begin{pmatrix}  \sum_{j \neq 1} \kappa_{1j} & -\kappa_{12}  & \dots & -\kappa_{10}\\
-\kappa_{21} &   \sum_{j \neq 2} \kappa_{2j} & \cdots  & -\kappa_{20}\\
\vdots & \vdots & \ddots & \vdots\\
-\kappa_{01} & -\kappa_{02} & \dots & \sum_{j \neq 0} \kappa_{0j} \end{pmatrix}, 
$$
 and hence has the same structure as $\Qv$.

  \subsection{Prior on the indicator variables for an ordinal covariate}\label{app:pridel_ord}
  For an ordinal covariate  the structure matrix $\Qv(\deltav)$ given in equation (\ref{eq:priQord}) can be written as the  product 
  $$\Qv(\deltav)=    \Dv' \kappav \Dv,$$
  where $\Dv$ is a first order  difference matrix  
  and $\kappav$ a diagonal matrix  with elements $\kappa_{k,k-1}$
    $$\Dv=
  \begin{pmatrix} 1 & 0 & 0 & \hdots &0 &0\\
                                              -1 & 1 & 0 & \hdots &0 & 0 \\
                                              \vdots&\vdots&\vdots&\ddots &\vdots &  \vdots\\
                                            0  & 0 & 0&\hdots & -1 & 1\end{pmatrix}  \qquad  \text{and} \quad 
                                           \kappav =\begin{pmatrix}   \kappa_{10} & 0 &0 & \hdots &0 \\
                                                                  0&  \kappa_{12} & 0 & \hdots& 0 \\
                                                                      \vdots&\vdots&\vdots&\ddots &\vdots\\
                                                                      0  & 0 & 0& \hdots &  \kappa_{c,c-1} \end{pmatrix}. $$
 As $|\Dv|=1$, the determinant of $\Qv(\deltav)$ is given as
  $$|\Qv(\deltav)|=\prod_{k=1}^c \kappa_{k,k-1}= r^{\sum_k (1-\delta_{k,k-1})},$$ and hence 
 $$p(\deltav)\propto |\Qv(\deltav)|^{-1/2}r^{\sum_k (1-\delta_{k,k-1})/2)} =1.$$

  In standard variable selection $\Qv(\deltav)=\diag(\kappa_{10},\dots,\kappa_{c0})$ and hence
  $$p(\deltav)\propto (\prod_{k=1}^c r^{1-\delta_{k0}})^{-1/2}r^{\sum_k (1-\delta_{k0})/2} =1.$$
   \section{Posterior inference}  \label{app:postinf}
   \subsection{Propriety of the posterior distribution}\label{app:prop}
  In our  specification only  the prior  on the error variance is improper if
  $p(\sigma^2 )\propto 1/\sigma^2$.  Propriety of the posterior distribution can be established even  for the improper prior $p(\mu,\sigma^2)=\frac{1}{\sigma^2}$ using a result of \cite{sun-etal:prop}. 
  With a slightly different notation than in their paper the regression model considered in  \cite{sun-etal:prop}  can be written as
  $$\yv=\Zv_\alpha \alphav+ \Zv_\beta \betav +\errorv \qquad \errorv \sim \Normal{0,\sigma^2 \mathbf{I}}$$ 
   where $\Zv_\alpha$ is the $n \times f$  regressor matrix  for  the  fixed effects $\alphav$ and $\Zv_\beta$ is the  $n \times q$ regressor matrix for the random effects $\betav=(\betav_1',\dots,\betav_p')'$, where   $\betav_h$ is  a column  vector  of dimension $c_h$, $h=1,\dots,p$.  $\betav$  follows a multivariate   Normal distribution with a blockdiagonal variance-covariance matrix $\diag(\tau^2_1\Bv_1,\dots, \tau^2_p\Bv_p)$ where the scale parameters $\tau^2_h$ are allowed to differ across blocks and are assigned Inverse Gamma priors $\tau^2_h \sim \Gammainv{g_{h0}, G_{h0}}$.
 
 This is  exactly the setting of our  regression model (\ref{regmod1}) where conditioning on the vector $\deltav$ the intercept is the only  fixed effect, i.e. $f=1$ and  the regression effects $\betav_h$ have independent  Normal priors   with  Inverse Gamma  hyperpriors on the scale parameters  $\tau^2_h$, and thus correspond to random  effects in \cite{sun-etal:prop}.
  
 Let   $$SSE=\yv' (\Iv-\Zv(\Zv'\Zv)^-\Zv')\yv$$
  where  $\Zv=(\Zv_\alpha, \Zv_\beta)$ and $(\Zv'\Zv)^-$ is a generalised inverse of  $\Zv'\Zv$.
     
Theorem 2 of \cite{sun-etal:prop} states, that  if $2 S_0 +SSE>0$, the following  conditions are 
     sufficient  for propriety of the posterior  under   a flat prior on $\alphav$:  
  \begin{enumerate} 
  \item[(a)]  $G_{h0}>0$   for $h=1,\dots, p$
  \item[(b)] $c_h + 2 g_{h0} > q-t$  for all $h=1,\dots,p$
 \item [(c)] $n-f+2 s_0>0$  
 \end{enumerate}
where $t$ is the rank of $\Xv_\beta'\big (\Iv-\Xv_\alpha(\Xv_\alpha' \Xv_\alpha)^{-1} \Xv_\alpha\big)\Xv_\beta$.
 As we use proper priors on the scale parameters $\tau^2_h$ with $g_{h0}>0$ and $G_{h0}>0$, propriety of 
 the posterior is guaranteed   if $SSE>0$ also for $s_0=S_0=0$  if $n>f$  and  condition (b) holds.   
 
In a model where the intercept is the only  fixed effect, $q=t$ if the design matrix $[\mathbf{1},\Xv_\beta]$ has full column rank $q+1$. Then 
$SSE>0$  with probability 1, $n\ge f=1$, and  condition (b) holds for all $h$.
 

   \subsection{Details on sampling steps}  \label{app:detinf}
    Here  we give details on the sampling steps of the MCMC scheme outlined in Section \ref{sec:MCMC}. To initialise the sampler we choose  starting values for the error variance $\sigma^2$ as well as   the vector of indicators $\deltav$ and the scale parameters $ \tauv^2$ 
    and compute the prior covariance matrix $\Bv_0(\deltav,\tauv^2)$. Then the following sampling steps are iterated: 
    \begin{enumerate}
    \item[(1)]   Sample the vector of regression coefficients $\betav$ from the full conditional $$p(\betav|\sigma^2, \deltav,\tauv^2,\yv )\propto p(\yv|\betav,\sigma^2)p(\betav|\deltav,\tauv^2),$$ which is  the Normal distribution, $\Normal{\bv,\Bv}$ with moments given as        
 \begin{align*}
		\Bv^{-1}& = \left(\Bv_0(\deltav,\tauv^2)\right)^{-1}+ \frac{1}{\sigma^2 }\Xv'\Xv \\ 
					\bv &=\frac{1}{\sigma^2} \Bv  \Xv' \yv.
		\end{align*}

     \item[(2)] Sample the  error variance $\sigma^2$  from its full conditional $p(\sigma^2| \betav, \deltav,\tauv^2, \yv)$,
       which  is   the Inverse 
	Gamma distribution $\Gammainv{s,S}$  with parameters 
	\begin{align*}
		s&=s_0+n/2 \\
		S&=S_0+\frac{1}{2}(\yv-\Xv \betav)'(\yv-\Xv\betav)
	\end{align*}

      \item[(3)] Sample the  scale parameters  $\tauv^2$  from the full conditional $p(\tauv^2| \betav, \sigma^2, \deltav,\yv)$.
      
      As the  full conditional for $\tauv^2$ is given as
       \begin{align*} p(\tauv^2| \betav, \sigma^2,\deltav,\yv) &  \propto \prod_{h=1}^p p(\tau_h^2)\cdot p(\betav_h|\deltav_h,\tau_h^2 )  =\\
       & \prod_{h=1}^p   (\tau^2_h)^{-g_{h0}-1} \exp(-\frac{G_{h0}}{ \tau_h^2}) \cdot (\tau^2_h)^{-c_h/2} \exp(-\frac{\betav_h' \Qv_h(\deltav_h)\betav_h}{2 \gamma_h \tau_h^2} )
       \end{align*}
 the scale parameters $\tau^2_h$, $h=1,\dots,p$ are independent a posteriori  and the full  conditional  for $\tau^2_h$ is  
  the    Inverse Gamma distribution $\Gammainv{g_h, G_h}$
	  with parameters 
		\begin{align*}
			g_h& = g_{h0} + \frac{c_h}{2}\\
			G_h& = G_{h0}+ \frac{1}{2\gamma_h }\betav_h' \Qv(\deltav_h)\betav_h. 			
		\end{align*}
	
	 \item[(4)]  For all $h \in \{1,\dots, p\}$ where  a hyperprior $G_{h0} \sim \Exp{\lambda_h}$ is specified $G_{h0}$ is sampled from  its full conditional
	 	  $$p(G_h|\tau_h^2)\propto p(G_h)p(\tau_h^2|G_h)= \exp(-\frac{G_h}{\lambda_h}) \cdot G_h^{g_h} \exp(-\frac{G_h}{\tau^2_h}), $$
which is the Gamma distribution $\Gammad{g_h+1, \frac{1}{\lambda_h}+\frac{1}{\tau^2_h}}$.

    \item[(5)] Sample the vector of binary indicators  $\deltav$  from the full conditional given as 
   \begin{align*} p(\deltav| \betav, \sigma^2, \tauv^2, \yv,) & \propto p(\betav|\deltav,\tau^2)p(\deltav)
   \propto \prod_{h=1}^k  \exp(-\betav_h'\Qv(\deltav_h)\betav_h) \sqrt{r}^{\sum(1-\delta_{h,kj})} =\\
   &=\prod_{h=1}^k \prod_{k\neq j}(\sqrt{r})^{1-\delta_{h,kj}}\exp\Big(-\frac{(\beta_{h,k}-\beta_{h,j})^2}{2\tau_h^2\gamma_h} \big(\delta_{h,kj}+r(1-\delta_{h,kj})\big)\Big)
   \end{align*}
   see equations (\ref{eq:postdel}) and (\ref{eq:postdel1}).

  Hence from equation (\ref{eq:condpri}) we get
    $$p(\delta_{h,kj}=1|\beta_{h,k}, \beta_{h,j} , \tau_h^2) = \frac{1}{1+L_{h,kj}},$$
		where
	$$ L_{h,kj}= \sqrt{r}\exp\Big(-\frac{r-1}{2} \frac{(\beta_{h,k}-\beta_{h,j})^2}{ \gamma_h\tau_h^2}\Big). $$
          
 \item[(6)]   For $h=1,\dots, p$ compute $\Bv_{h0}(\deltav_h, \tauv_h^2))= \gamma_h \tau^2_h \Qv^{-1}(\delta_h)$ and 
 update the prior covariance matrix $\Bv_0(\deltav, \tauv^2)$.
   \end{enumerate}
   
   \subsection{Computation times}\label{app:comptime}
   Table  \ref{app:tab_comptimes} shows  the computation time (in sec.) needed for $m=1000$ MCMC steps of effect fusion in a  regression model  with one nominal covariate with $c$ levels (except baseline)
  for $n$  observations. Computations were performed on a laptop with  Intel Core i7-5600U processor with 2.60 GHz and 16 GB RAM.
  \begin{table}[!ht]
	\centering
		\begin{tabular}{r c  |  r c}
  \multicolumn{2}{c|}{n=1000}  & \multicolumn{2}{c}{n=10000}\\
     c & time & c & time\\ \hline
     10 &0.4 & 10 & 0.6 \\
   20 & 1.1 &  20 & 1.5 \\
    50 &45.3 & 50 & 47.8  \\
    100 &1536.6& 100 & 1539.7\\
       \end{tabular}
       \caption{Computation (in sec.) times for $n$ observations and one nominal covariate with $c$ levels (except baseline).  } 
    \label{app:tab_comptimes}
  \end{table}

  \section{Details on  the simulation study} \label{app:sim}
  
  \subsection{Alternative methods}\label{app:sim_meth}
  
  Table \ref{app:tab_methods}  lists the  methods  to which we compare Bayesian effect fusion in the simulation study,
  together with the name of the corresponding {\tt R} package and  references. The code of the Graph Laplacian approach in \cite{liu-etal:bay} was provided  directly from the authors. 
  
  \begin{table}[ht]
  \centering
  \small
  \begin{tabular}{lll}
  	Method & {\tt R} package & References \\
  	\hline
  	Penalty & gvcm.cat & \cite{ger-tut:spa} \\
  	BLasso & monomvn & \cite{par-cas:bay} \\
  	BEN & EBglmNet & \cite{hua-etal:emp} \\
  	GLasso & grpreg & \cite{yua-lin:mod} \\
  	GLap & - & \cite{liu-etal:bay} \\
  	SGL & SGL & \cite{sim-etal:spa}\\
  	BSGS & BSGS & \cite{che-etal:bay}, \cite{lee-chen:bay}
  \end{tabular}
    \caption{Details for methods used in the simulation study} 
    \label{app:tab_methods}
  \end{table}

  \subsection{Evaluation of model selection} \label{app:sim_measures}
  
  To evaluate   model selection, we use  the  true positive rate (TPR), the true negative rate (TNR), the positive predictive value (PPV) and  the negative predictive value (NPV). These measures  are defined as 
  \begin{align*}
  	\text{TPR} & =\text{ TP} / \text{P } \cdot 100\%= \text{TP} / (\text{TP} + \text{FN)} \cdot 100\%, \\
  	\text{TNR} & = \text{TN} / \text{N}  \cdot 100\% = \text{TN }/ (\text{TN} + \text{FP})  \cdot 100\%,\\
  	\text{PPV} & = \text{TP} / (\text{TP }+\text{ FP)}   \cdot 100\%,\\
  	\text{NPV} & = \text{TN} / (\text{TN} +\text{ FN})  \cdot 100\%,
  \end{align*}
 where TP (\emph{true positive}) is the number of correctly detected non-zero effect differences, TN (\emph{true negative})  the number of correctly  detected zero differences, FN (\emph{false negative}) the number of zero effect differences  classified as non-zero and FP (\emph{false positive})  the number of  zero effect differences classified as non-zero.

\end{document}